\documentclass[sigconf, nonacm, pdfa]{acmart}

\newcommand\vldbdoi{10.14778/3749646.3749667}
\newcommand\vldbpages{3965 - 3978}
\newcommand\vldbvolume{18}
\newcommand\vldbissue{11}
\newcommand\vldbyear{2025}
\newcommand\vldbauthors{\authors}
\newcommand\vldbtitle{\shorttitle} 
\newcommand\vldbavailabilityurl{https://github.com/ah89/DobLIX}
\newcommand\vldbpagestyle{empty} 

\usepackage{amsmath,amsfonts}
\usepackage{textcomp}
\usepackage[linesnumbered,ruled,vlined]{algorithm2e}
\usepackage[percent]{overpic}
\usepackage{algpseudocode}
\usepackage{MnSymbol}
\usepackage{wasysym}
\usepackage{multirow}
\usepackage{subcaption}
\usepackage{enumitem}
\usepackage{graphicx} 
\usepackage{tikz}
\usepackage{pdfx}
\usepackage{balance}
\newcommand*\circled[1]{\tikz[baseline=(char.base)]{
            \node[shape=circle,draw,inner sep=0.5pt] (char) {#1};}}

\usepackage[utf8]{inputenc}
\usepackage[T1]{fontenc}
\usepackage{microtype}
\usepackage{graphicx}
\usepackage{balance}  %

\usepackage{multirow}
\usepackage{color}
\usepackage{listings}
\usepackage{float}
\usepackage{outlines}
\usepackage[normalem]{ulem}
\usepackage{cleveref}
\usepackage{enumitem}
\usepackage{soul}
\usepackage{cuted, lipsum}
\usepackage[listings,skins,breakable]{tcolorbox}
\usepackage{booktabs}
\PassOptionsToPackage{rgb,hyperref,table}{xcolor}
\usepackage{xcolor}

\usepackage{colortbl}
\usepackage{marginnote}
\usepackage{dirtytalk}
\usepackage{etoolbox}
\BeforeBeginEnvironment{example}{\vspace{-0.7em}}
\AfterEndEnvironment{example}{\vspace{-0.7em}}

\usepackage[skip=0pt]{caption} 

\renewcommand{\abstractname}{ABSTRACT}

\usepackage{titlesec}

\titlespacing*{\section}{1pt}{2pt}{3pt}         
\titlespacing*{\subsection}{1pt}{2pt}{3pt}
\titlespacing*{\subsubsection}{0pt}{1pt}{2pt}
\titlespacing*{\example}{0pt}{0pt}{2pt}

\begin{document}
\title{\texttt{DobLIX}: A Dual-Objective Learned Index for Log-Structured Merge Trees}

\author{Alireza Heidari}
\affiliation{%
  \institution{Huawei}
}
\email{alireza.heidarikhazaei@huawei.com}

\author{Amirhossein Ahmadi}
\affiliation{%
  \institution{Huawei}
}
\email{amirhossein.ahmadi@huawei.com}

\author{Wei Zhang}
\affiliation{%
  \institution{Huawei}
}
\email{wei.zhang6@huawei.com}

\begin{abstract}
In this paper, we introduce \texttt{DobLIX}, a dual-objective learned index (LI) specifically designed for Log-Structured Merge (LSM) tree-based key-value stores. Traditional LIs primarily focus on optimizing index lookups, often overlooking the critical role of data access from storage, which can become a significant performance bottleneck. In LSM-based systems, a considerable portion of the index is stored on disk, making lookups highly dependent on the efficient coordination between in-memory structures and disk-resident data. Poorly optimized access patterns can lead to excessive I/O operations, negatively impacting read latency and overall system performance. \texttt{DobLIX} addresses this by incorporating a second objective, data access optimization, into the LI training process. This dual-objective approach ensures that both index lookup efficiency and data access costs are minimized, leading to significant improvements in read performance while maintaining write efficiency in real-world LSM systems. Additionally, \texttt{DobLIX} features a reinforcement learning agent that dynamically tunes the system parameters, allowing it to adapt to varying workloads in real-time. Experimental results using real-world datasets demonstrate that \texttt{DobLIX} reduces indexing overhead and improves throughput by $1.19\times$ to $2.21\times$ compared to state-of-the-art methods within RocksDB, a widely used LSM-based storage engine.
\end{abstract}

\maketitle

\pagestyle{\vldbpagestyle}
\begingroup\small\noindent\raggedright\textbf{PVLDB Reference Format:}\\
\vldbauthors. \vldbtitle. PVLDB, \vldbvolume(\vldbissue): \vldbpages, \vldbyear.\\
\href{https://doi.org/\vldbdoi}{doi:\vldbdoi}
\endgroup
\begingroup
\renewcommand\thefootnote{}\footnote{\noindent
This work is licensed under the Creative Commons BY-NC-ND 4.0 International License. Visit \url{https://creativecommons.org/licenses/by-nc-nd/4.0/} to view a copy of this license. For any use beyond those covered by this license, obtain permission by emailing \href{mailto:info@vldb.org}{info@vldb.org}. Copyright is held by the owner/author(s). Publication rights licensed to the VLDB Endowment. \\
\raggedright Proceedings of the VLDB Endowment, Vol. \vldbvolume, No. \vldbissue\ %
ISSN 2150-8097. \\
\href{https://doi.org/\vldbdoi}{doi:\vldbdoi} \\
}\addtocounter{footnote}{-1}\endgroup

\ifdefempty{\vldbavailabilityurl}{}{
\vspace{.3cm}
\begingroup\small\noindent\raggedright\textbf{PVLDB Artifact Availability:}\\
The source code, data, and/or other artifacts have been made available at \url{\vldbavailabilityurl}.
\endgroup
}
\section{INTRODUCTION}
\label{sec:intro}
\noindent
\textbf{Context.} Key-value (KV) databases are crucial in various domains like cloud computing, e-commerce, big data analysis, and artificial intelligence. Among different KV store architectures, Log-Structured Merge Trees (LSMs) stand out for their exceptional write performance~\cite{CliffGuard}. They are widely used in industrial applications, such as 
RocksDB~\cite{rocksdbpaper}, LevelDB~\cite{leveldb}, Cassandra~\cite{lakshman2010cassandra}, and BigTable~\cite{chang2008bigtable}.

\begin{figure}[t]
  \centering
  \makebox[\columnwidth][c]{\includegraphics[width=\columnwidth]{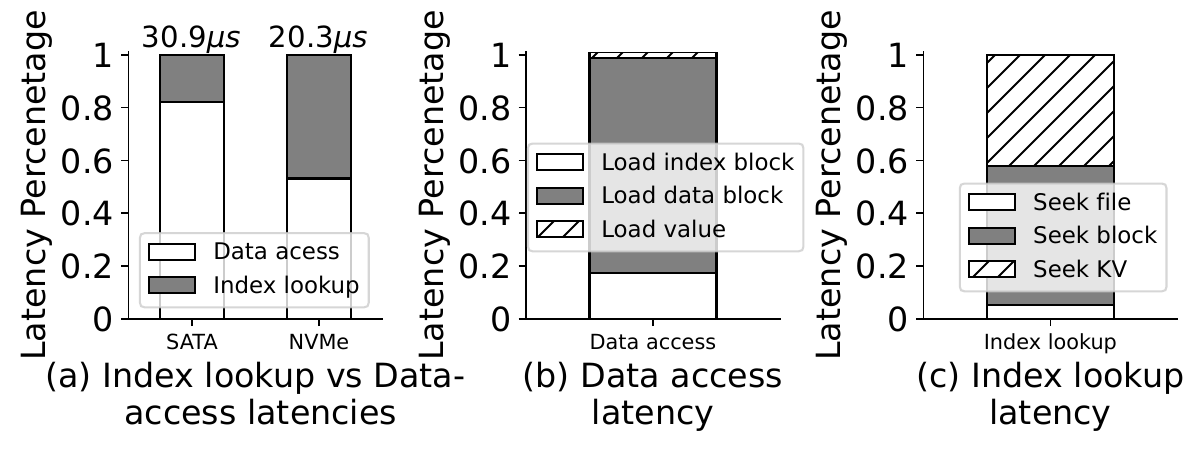}}
  \caption{\small Lookup Latency Breakdown. Read performance on $10$ million $8$-byte KVs of the Wiki dataset using the native RocksDB index.}
  \label{fig:rocksdb-lookup}
\end{figure}

The multi-level structure of LSMs~\cite{mo2025grow,sarkar2023lsm} results in significant drops in read performance due to high read amplification~\cite{TridentKV2022}. The levels are divided into Sorted String Tables (SSTs) that contain KV pairs ordered by keys, with SSTs comprising fixed-size blocks (ranging from $4KB$ to $32KB$), and retrieving a specific KV pair involves an index lookup to locate the relevant level, file, block, and pair, followed by a data-access phase to load the block and fetch the desired pair. In current KV stores, with modern NVMe storage, both index lookup and data access contribute significantly to query latency (Fig.~\hyperref[fig:rocksdb-lookup]{\ref*{fig:rocksdb-lookup}a}).

A common strategy to enhance index lookup is Learned Indexing (LI), which involves using Machine Learning (ML) as a data structure, effectively revolutionizing data indexing by leveraging predictive capabilities and efficiently modeling records distribution patterns~\cite{ding2020alex,heidarikhazaei2021structured,ferragina2020pgm,galakatos2019fiting,heidari2019holodetect,heidari2020sampling}. These LI models usually reside inside memory and estimate the probable range containing a specific record by evaluating the target key through a monotonic function approximated by a cumulative distribution function (CDF)~\cite{case2023memory,heidari2020approximate}. However, when KV pairs are stored in storage, the LI bottleneck emerges during the data retrieval process from storage to memory~\cite{lidisk2024sigmod,wipe22024,apex2021,liDisk2024}
($\S\ref{sec:li-storage})$. Consequently, to enhance performance overall, it is essential to take into account these effective factors. 

\noindent

\noindent
\textbf{Current LI on LSM Research.}
The potential enhancement of LSM lookup performance through LI integration has been a focal point in recent studies such as \textit{TridentKV}~\cite{TridentKV2022}, \textit{Bourbon}~\cite{Bourbon2020}, \textit{LeaderKV}~\cite{leaderkv2024}, and \textit{LearnedKV}~\cite{wang2024learnedkv}. These initiatives have introduced customized LI solutions for LSM structures, showing performance gains in \textit{index lookup efficiency}. However, a \textbf{critical yet overlooked challenge} is that \textit{indexing in LSMs is not merely about predicting data locations—it must also ensure efficient data access}. Unlike traditional small data structures, where indexes fit entirely in memory, LSM indexes often span both \textbf{memory and disk}, requiring frequent storage accesses. Consequently, \textbf{optimizing the index without considering how data is retrieved from storage can lead to inefficiencies}, such as excessive block I/O or misaligned lookups~\cite{heidari2024metahive}. Fig.~\hyperref[fig:rocksdb-lookup]{\ref*{fig:rocksdb-lookup}b} shows that the retrieval of blocks from storage to memory is a dominant factor in lookup latency.  

\begin{figure}
  \centering
  \makebox[\columnwidth][c]{\includegraphics[width=\columnwidth]{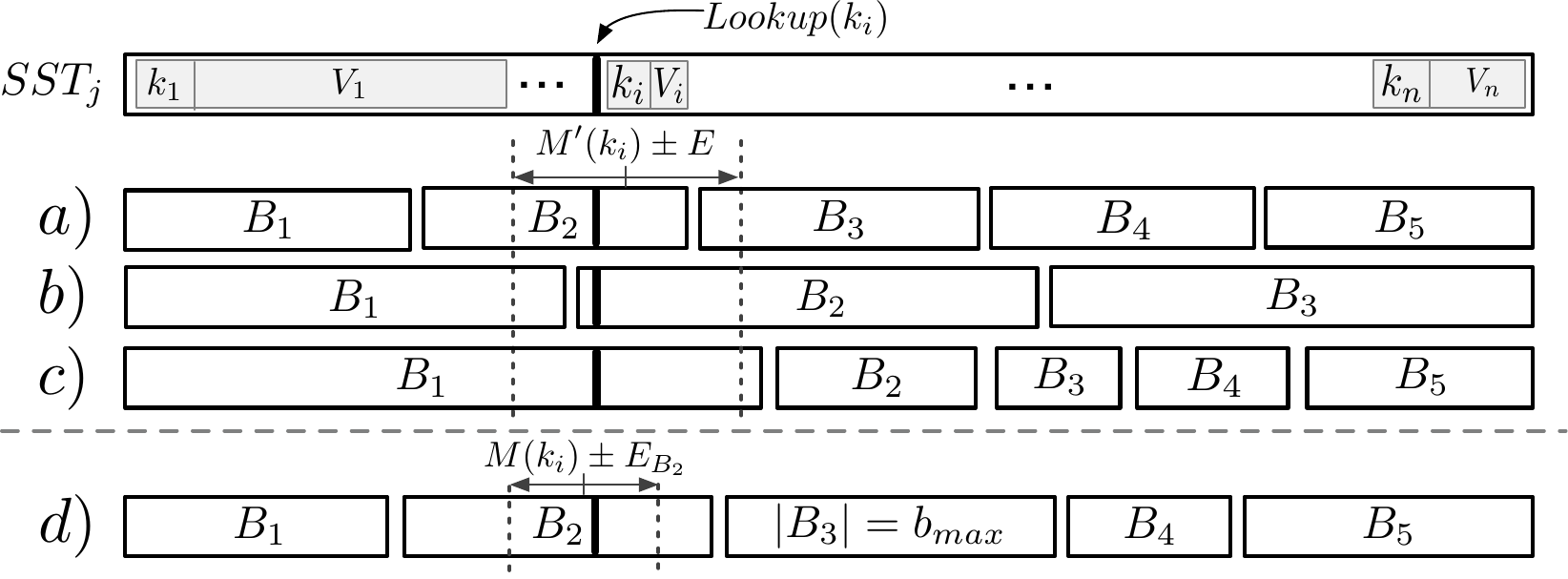}}
  \caption{\small{Comparison of LI Solutions on SSTs. Considering block partitioning $Par_{Block}$ and the indexing $I_{IndexBlock}$ as stochastic variables. \textbf{a)} Small fixed-size blocks, \textbf{b)} Large fixed-size blocks with a guarantee on max block size. \textbf{c)} Variable block size with a model output ($M'(k)$) guarantee to load one block. \textbf{d)} Perfect solution with guarantees on model output ($M(k)$) for one block access with optimized block size.}}
  \label{fig:prev-designs}
\end{figure}

This oversight results in a dilemma depicted in Fig.~\ref{fig:prev-designs} when previous methodologies neglect this characteristic:

\begin{enumerate}[noitemsep, leftmargin=*, topsep=2pt, partopsep=0pt, itemsep=2pt, parsep=0pt]
\item \textit{Adhering to fixed block sizes optimized for read operations}, which can result in inaccuracies in block lookup and require accessing multiple blocks during read processes. Solutions \textit{a} and \textit{b}, as depicted, face this challenge, a methodology also used by \textit{Bourbon} and \textit{LearnedKV}~\cite{wang2024learnedkv}.

\item \textit{Opting for variable block sizes}, which can lead to significantly larger blocks, thus increasing block I/O time. Solution \textit{c} exhibits this issue, a methodology also adopted by \textit{TridentKV}~\cite{TridentKV2022} and \textit{LeaderKV}~\cite{leaderkv2024}.
\end{enumerate}

These dilemmas arise from the \textbf{isolated optimization of LI models without considering the data access component}. Consequently, while these techniques reduce average latency compared to LSM-based KV store index lookups, they show \textbf{higher tail-latency} in cases requiring multiple or large block retrievals. This paper proposes a new LI solution to optimize \textbf{both the indexing and data access phases}, ensuring that predicted locations align with an efficient storage layout.  

Furthermore, earlier studies have neglected the consideration of varying key and values lengths in their design (\S\ref{sec:var-kv-lsm}), as well as the enhancement of the \textbf{last-mile search}, which is the final phase to identify the target KV within the retrieved block. As shown in Fig.~\hyperref[fig:rocksdb-lookup]{\ref*{fig:rocksdb-lookup}c}, this step in the indexing process incurs nearly identical latency to block finding. \textbf{Thus, an LI methodology should integrate optimizations that jointly improve index lookup, block access, and last-mile search efficiency.}

\noindent\textbf{Our Approach.} We leverage multi-objective optimization techniques to incorporate data access overhead as an additional factor to design an LI framework~\cite{boyd2004convex}. To achieve a suitable configuration that optimizes index performance alongside this additional parameter, we employ an auto-tuning method based on reinforcement learning (RL). This enables us to develop configurations that generate index models capable of automatically adjusting when workloads change. Our approach specifically concentrates on constructing an index model for KVs at the SST level, leveraging their inherent sorted structure for efficient LI training. The resulting model enables \textbf{effective locating} of the desired \textbf{variable-size} KVs using merely \textbf{a single block} access with \textbf{ideal block size}, all while maintaining the LI model optimal performance.

We evaluate our framework using an array of real-world and synthetic datasets and workloads, showing a substantial reduction in indexing costs when measured against state-of-the-art indexing solutions. We show that our framework leads to better performance, with throughput improvements between $1.19\times$ and $2.21\times$ in various datasets and workloads. 

\noindent\textbf{Technical Challenges.} Our approach needs to address multiple technical challenges:  

\begin{itemize}[noitemsep, leftmargin=*, topsep=2pt, partopsep=0pt, itemsep=2pt, parsep=0pt]
    \item \textit{\textbf{Index Model.}} A second parameter complicates index modeling. The model must optimize both parameters, but the cost feedback for the second one may be asynchronous.
    \item  \textit{\textbf{Restoration.}} During LSM query processing, data spans storage layers with varying performance. Missing metadata (e.g., the \say{index trajectory}) can disrupt queries and require costly recovery. To avoid this, index system must preserve critical metadata to efficiently handle gaps and ensure smooth query execution.
    \item  \textit{\textbf{Adaptation.}} In an index framework, the parameters of the system can change due to the pattern of the incoming data or workload shifts. The design must adapt to these changes to ensure efficient performance under varying conditions.
\end{itemize}

\noindent Implementing the framework within an LSM-based KV store and conducting comprehensive benchmarking present challenges, primarily due to the incompatibility of many LI models with LSMs, as well as the inadequacies of current benchmarks (e.g., YCSB) in capturing important edge cases -- including variable KV lengths, compaction overheads, and granular lookup latency breakdowns.

\noindent
\textbf{Contributions.} We present the following contributions in this paper:
\begin{enumerate}[noitemsep, leftmargin=*, topsep=2pt, partopsep=0pt, itemsep=2pt, parsep=0pt]
    \item \textbf{\textit{Proposing \texttt{DobLIX}, a \underline{D}ual-\underline{ob}jective \underline{L}earned \underline{I}nde\underline{X} framework}}, for LSM-based KV stores ($\S\ref{sec:design:overview}$). This framework optimizes the performance of the LI lookup while considering any additional secondary objective parameter using two innovative LI approximation models, PLA and PRA ($\S\ref{sec:design:li}$). In this work, our design specifically optimizes indexing and data access as a secondary parameter.
    \item \textbf{\textit{Optimizing the last-mile search phase}} of the lookup process by incorporating the LI model traversal into the last mile search process ($\S\ref{sec:design:lastmile}$).
    \item \textbf{\textit{Introducing an RL-based agent}} to dynamically adjust the indexing and data partitioning parameters in our system and to choose between the PLA and PRA methods ($\S\ref{sec:design:agent}$).
    \item \textbf{\textit{Implementing our method on RocksDB}} and performing a comprehensive benchmark to demonstrate its superior performance compared to traditional and LI solutions ($\S\ref{sec:eval}$).
\end{enumerate}

\noindent\textbf{Paper Organization.} The remainder of the paper proceeds as follows: In \S\ref{sec:background}, we review the background concepts. \S\ref{sec:doblix-design} provides a detailed overview of our LI framework design and RL-agent. In \S\ref{sec:eval}, we evaluate our proposed solutions. We highlight related work in \S\ref{sec:related-work}. We discuss our method portability and tuning agent robustness in \S\ref{sec:discussion}, and conclude the key points of the paper and future directions in \S\ref{sec:conclusion}.

\section{BACKGROUND}
\label{sec:background}

\subsection{Data Management and LSM-based KV Stores}
\label{sec:rocksdb}
To formally define the data management system's random variables in Fig.~\ref{fig:prev-designs}, we must express them mathematically~\cite{2024limousine}. First, we know two main ways to simplify big data management complexity: (1) \textit{Data Partitioning ($P(.)$)}: splits data into smaller parts~\cite{sarkar2022dissecting}. (2) \textit{Indexing ($I(.,.)$)}: returns partition(s) containing the query key.

\noindent$P(.)$ is unchanged by the key query but may integrate with $I_{LSM}(.,.)$ for given key $k$. Consequently, for simplicity in notation during the construction phase, we can represent indexing as $I(.)$. This approach enables creating indexed data structures for any dataset $D$. A general combination $\mathcal{F}=f_1\circ f_2\circ\dots\circ f_k(D)$, where each $f_i$ is from a set of partitions $\mathcal{P}=\{P_1,\dots,P_n\}$, or a set of indexes $\mathcal{I}=\{I_1,\dots,I_m\}$ forms a data structure. Some structures may be infeasible since partitions $P_i$ or indices $I_i$ assume specific input data properties. After construction, $\mathcal{F}(k)$ becomes the data structure's top-level index.

In the context of an LSM, the data structure is constructed from four distinct partitions defined as $$\mathcal{P}_{LSM}=\{ P_{TreeLevel}, P_{SST}, P_{Block}, P_{KV} \}.$$
Its indexing mechanism is represented by $$\mathcal{I}_{LSM}=\{ I_{LevelBloomFilter}, I_{SST}, I_{IndexBlock}, I_{KV} \}.$$
The composite function is articulated as 
\begin{equation}
    \begin{split}
    \mathcal{F}_{LSM} =& I_{KV} \circ P_{KV}\circ I_{IndexBlock}\circ P_{Block}\circ I_{SST}\circ \\ &P_{SST} \circ I_{LevelBloomFilter}\circ P_{TreeLevel}(D).
\end{split}
\end{equation}

\noindent $\mathcal{F}_{LSM}$ partitions data $D$ hierarchically via tree levels ($P_{TreeLevel}$), using a level bloom filter ($I_{LevelBloomFilter}$) that marks key presence at each level~\cite{dayan2018optimal}. Each level stores data in separate SST files ($P_{SST}$) with a linear-search index ($I_{SST}$) tracking key ranges. Each SST sorts data and splits it into blocks ($P_{Block}$) with key-value pairs ($P_{KV}$). Each SST’s metadata has an index block with entries ($I_{IndexBlock}$) for data blocks enabling binary search, and KV pairs ($I_{KV}$) accessed via linear search in RocksDB. Fig.~\ref{fig:prev-designs} shows four $P_{Block}$ setups \{a,b,c,d\} that each yield different block divisions in one SST file; see \S\ref{sec:design:overview} for details.

\begin{figure}[t]
  \centering
  \makebox{\includegraphics[width=\columnwidth]{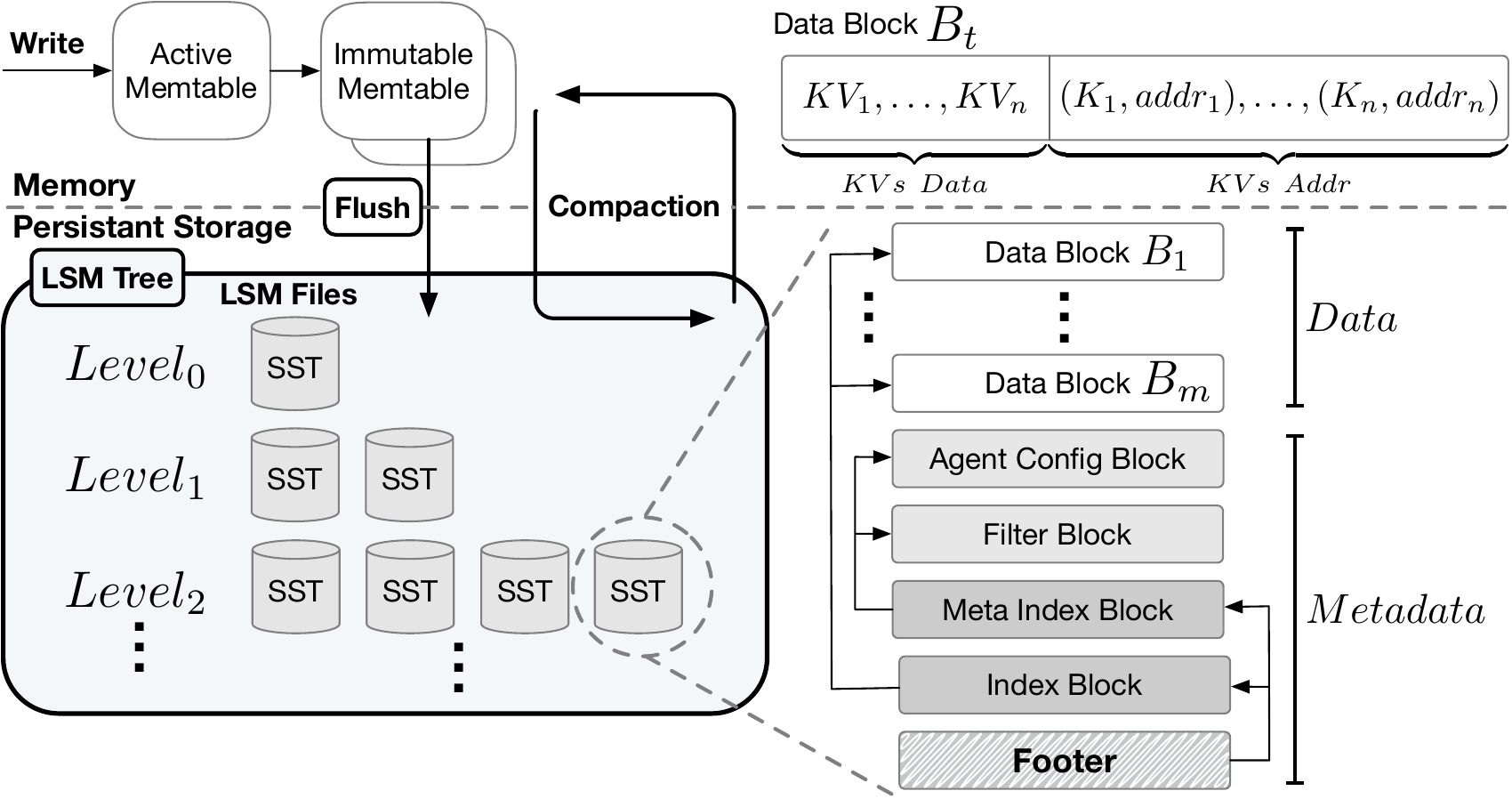}}
  \caption{\small{RocksDB Architecture.}}
  \label{fig:sst-format}
\end{figure}

\subsection{Learned Index (LI)}
\label{sec:li-storage}

LIs~\cite{ding2020alex,ferragina2020pgm,li2020lisa,kipf2020radixspline,heidari2024uplif} use machine learning to efficiently map sorted keys to locations in databases. These indexes may use complex models like neural networks or simpler hierarchical linear models.
Traditional LIs use ensemble learning and hierarchical model structuring. The index model $I(k)$ uses $I(k) = M(k) \times N$, where $M$ is the CDF estimating $p(x \le k)$ and $N$ is the key count, to guide queries from the root to the correct layer. Due to complex models, many LIs use piecewise linear models to approximate the CDF~\cite{wongkham2022updatable,livshits2020approximate}. Predict the key's position as $pos = m \times k + a$ (with error $E$); $m$, $a$ are learned parameters, and $E$ crucial for final target key search. These indexes are memory efficient due to lightweight parameters like intercept and slope versus conventional indexes.

\noindent
\subsubsection{\textbf{LI on Persistent Storage.}}

\begin{figure}[t]
  \centering
  \makebox[\columnwidth][c]{\includegraphics[width=\columnwidth]{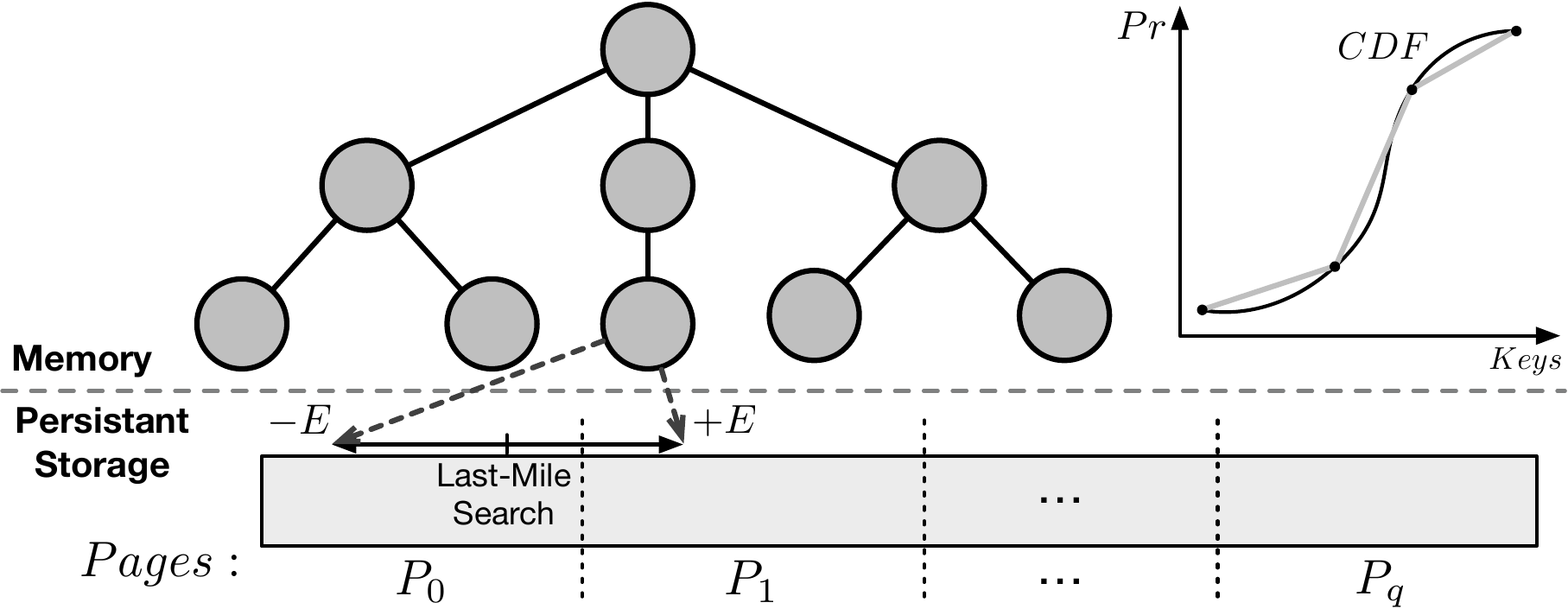}}
  \caption{\small{LI on Persistent Storage. The model last-mile search range may require loading multiple pages from the storage.}}
  \label{fig:li-storage}
\end{figure}

Studies~\cite{lidisk2024sigmod,wipe22024,apex2021,liDisk2024} show LIs fail to outperform B+tree for persistent storage KVs. These studies highlight that the design of LIs fails to leverage the characteristics of disk storage, necessitating multiple disk I/O operations during the last-mile search process as shown in Fig.~\ref{fig:li-storage}. A similar trend is observed in LI techniques for LSMs, which also entail substantial I/O operations during their final search phase. This underscores the importance of considering I/O as a critical element in the development of LI systems.

\noindent
\subsubsection{\textbf{LI on Strings.}}
\label{sec:li-on-strings}
Most LI research focuses on fixed-size keys, with few studies on variable-length strings \cite{rss,wang2020sindex,lits2024vldb}. The Radix String Spline (RSS) \cite{rss} uses a Trie-based method. In RSS, each Trie node processes $8$- or $16$-byte keys and builds an RS model for them. Memory architecture and language usually limit integer size. In 64-bit C++, the largest built-in integer is \texttt{\_\_uint64\_t}. Some compilers offer 128-bit integers (e.g. \texttt{\_\_uint128\_t}), enabling efficient conversion of Trie nodes from strings to integers. The RS model gives monotonic CDF predictions within a set error limit. If a key exceeds the error limit, RSS adds it to a redirector map and creates a new child node. RSS compares $8$- or $16$-byte key segments, saving costs over full-key checks, but last-mile search is still expensive. A study \cite{lits2024vldb} shows RSS spends over $70\%$ of its time on last-mile search, suggesting major room for optimizing LIs with string keys.

\subsection{Key-Value Length in LSMs}
\label{sec:var-kv-lsm}
LSM storage engines are commonly used in various applications. Thus, they must handle keys and values of any type or length. The study by Cao et al. \cite{modelrocksdb2020} examines the use of RocksDB in three different use cases at \textit{Meta}~(see Table~\ref{tab:kv-avg-sd}): \textbf{UDB} (storage engine of a SQL database), \textbf{ZippyDB} (storage engine of a distributed KV-store), and \textbf{UP2X} (persistent storage of an AI/ML service). The study presented the average (AVG) and standard deviation (SD) of the KV length in these applications. Previous LSM LI solutions~\cite{Bourbon2020,TridentKV2022,leaderkv2024,wang2024learnedkv} are limited to fixed key sizes, rendering them unsuitable for these applications.

\begin{table}[]
\centering
\caption{\small{AVG and SD of key-values (in bytes) on RocksDB use cases on UDB, ZippyDB, and UP2X.}}
\label{tab:kv-avg-sd}
\resizebox{0.6\columnwidth}{!}{%
\begin{tabular}{c|cc|cc}
\hline
\multirow{2}{*}{\textbf{Application}} & \multicolumn{2}{c|}{\textbf{Key}}               & \multicolumn{2}{c}{\textbf{Value}}              \\ \cline{2-5} 
                                      & \multicolumn{1}{c|}{\textbf{AVG}} & \textbf{SD} & \multicolumn{1}{c|}{\textbf{AVG}} & \textbf{SD} \\ \hline
\textbf{UDB}     & \multicolumn{1}{c|}{27.1} & 2.6 & \multicolumn{1}{c|}{126.7} & 22.1 \\ \hline
\textbf{ZippyDB} & \multicolumn{1}{c|}{47.9} & 3.7 & \multicolumn{1}{c|}{42.9}  & 26.1 \\ \hline
\textbf{UP2X}    & \multicolumn{1}{c|}{10.4} & 1.4 & \multicolumn{1}{c|}{46.8}  & 11.6 \\ \hline
\end{tabular}

}

\end{table}

\begin{figure*}
    \centering
        \begin{minipage}{0.515\textwidth}
        \centering
        \includegraphics[width=\textwidth]{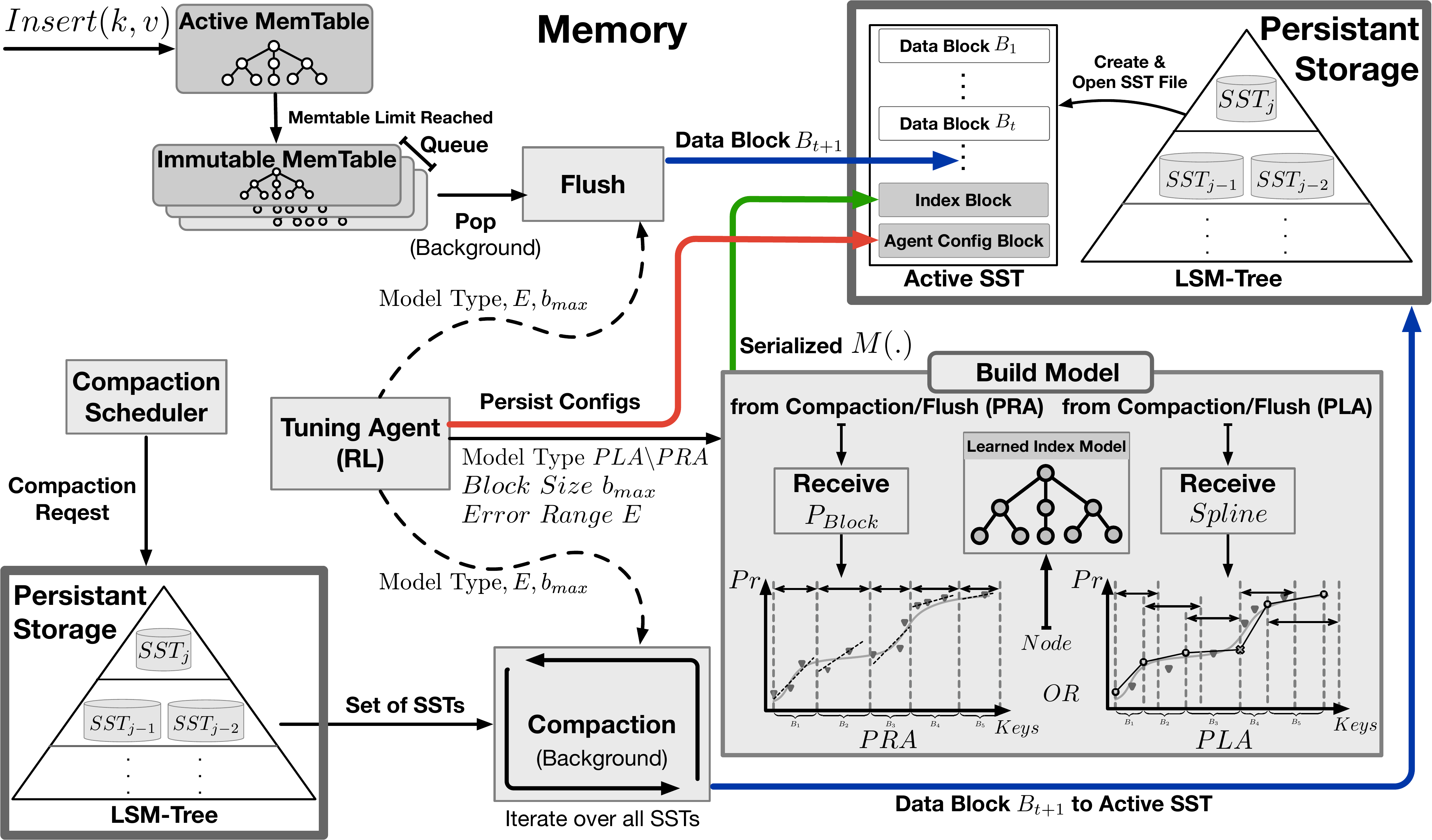}
        \captionof{figure}{{\small{\texttt{DobLIX} Architecture in Write Flow.}}}
        \label{fig:DobLIX-write}
    \end{minipage}
    \hfill
    \begin{minipage}{0.465\textwidth}
        \centering
        \includegraphics[width=\textwidth]{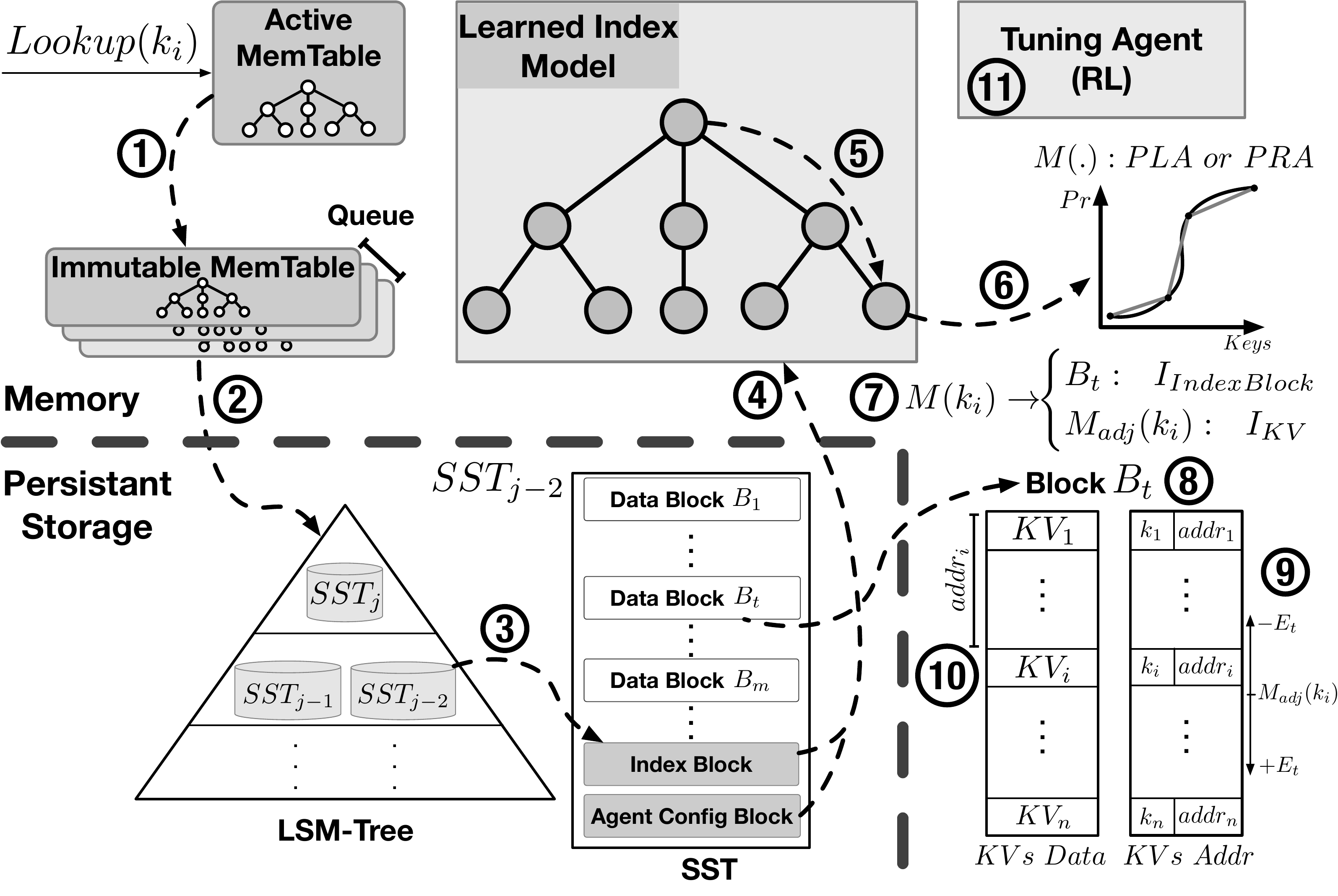}
        \captionof{figure}{\small{\texttt{DobLIX} Architecture in Read Flow.}}
        \label{fig:DobLIX-arch}
    \end{minipage}
\end{figure*}

\subsection{Lexicographic Optimization}
\label{sec:lexi_opt}
Lexicographic optimization~\cite{abernethy2024lexicographic,chinchuluun2007survey} is a multiobjective optimization approach that prioritizes objectives sequentially based on their importance. The method first optimizes the highest-priority objective, then optimizes subsequent objectives within the progressively constrained feasible region. The optimization process follows a strict hierarchy: first, optimize $f_1$ to find its optimum $f_1^*$. Next, optimize $f_2$ subject to $f_1(x) = f_1^*$, and continue similarly for subsequent objectives. This guarantees that higher-priority objectives remain uncompromised by lower-priority ones.
\begin{example}
Consider a resource allocation problem with three objectives ranked by importance: (1) minimize cost, (2) maximize quality, and (3) minimize delivery time. Using lexicographic optimization, we first find the solution with the lowest cost. Among these lowest-cost solutions, we select the one with the highest quality. If there are still ties, we choose the solution with the shortest delivery time. This method is useful when objectives must be addressed in a strict order of priority. In our method, I/O has more priority over index lookup performance.
\end{example}

\section{\texttt{DOBLIX} DESIGN}
\label{sec:doblix-design}

In this section, we describe how \texttt{DobLIX} is designed to speed up lookup queries. We begin by outlining \texttt{DobLIX}'s architecture and main ideas (\S\ref{sec:design:arch}, \S\ref{sec:design:overview}). To meet dual objectives, \texttt{DobLIX} uses two LI methods: The Piecewise Linear Approximation (PLA) method modifies spline definition rules to consider both objectives, while Piecewise Regression Approximation (PRA) improves performance by effectively managing modeling errors (§~\ref{sec:design:li}). \texttt{DobLIX} incorporates a string-compatible LI solution capable of handling variable-size KVs. To optimize the last-mile search process, it transfers model knowledge to narrow the search range and simplifies key comparisons by focusing solely on a limited part of the key bits decodable as an integer (\S\ref{sec:design:lastmile}). Furthermore, it uses an RL agent to adjust parameters like max error and block size, and to choose between PLA and PRA (\S\ref{sec:design:agent}).

\subsection{Overall Architecture}
\label{sec:design:arch}
Fig.~\ref{fig:DobLIX-arch} and \ref{fig:DobLIX-write} outlines the general architecture of \texttt{DobLIX}. This solution concentrates on learning the index at the SST level in detail. SSTs are preferred for LIs because their sorted and immutable KV structure eliminates the need for updates during their lifespan. Upon the formation of each SST, DobLIX trains an LI model based on its KVs. This model is crafted to accurately pinpoint the target block for all KVs within a specified error margin while ensuring that block sizes remain within the designated maximum limit. Consequently, \texttt{DobLIX} has an LI model and block partitioning, as illustrated in Fig.~\hyperref[fig:prev-designs]{\ref*{fig:prev-designs}d}. Subsequently, \texttt{DobLIX} serializes the LI model and deposits it in the index block within the SST metadata.
\noindent

{\small\textbf{Write Process.}} Fig.~\ref{fig:DobLIX-write} shows that SST creation and processing occur in the background, triggered either by a Flush to clear in-memory data or by Compaction scheduled for SST file maintenance in the LSM tree. Insert queries typically initiate the Flush process in the background. However, if the MemTable and queue are full, users may experience direct delays, as analyzed in our tail latency study (\S\ref{sec:eval:tail}). Flush/Compaction receives the latest parameters—maximum error $E$ and block size $b_{max}$—from the Tuning Agent for the current active SST and persists them in the Agent Config Block within the active SST for use during the lookup phase. Depending on the Model Type, PRA uses only $b_{max}$ to create block partitioning $P_{Block}$ and sends it to the Build Model module. If PLA is selected, it constructs block partitioning $Spline$ by $Spline$ and sends them to the module. The Build Model module inserts these signals to an optimized log-structured model as nodes for the lookup process described in \S\ref{sec:design:lastmile}, and finally stores the serialized model in the active SST Index Block.

\noindent
{\small\textbf{Lookup Process.}}
In Fig.~\ref{fig:DobLIX-arch}, the stages involved in a \texttt{DobLIX} lookup query are outlined. \circled{1} Initially, it inspects the current MemTables; if the desired key is absent there, it looks through the unalterable MemTables. \circled{2} It then scrutinizes various levels of LSMs\cite{dayan2018dostoevsky} and \circled{3} loads the SST that may cover the target key within its range. \circled{4} \texttt{DobLIX} loads the LI model from the SST metadata into memory. \circled{5} It performs a search within its Trie tree to locate the node that houses the ultimate LI (\S\ref{sec:li-on-strings}) and \circled{6} uses the trained CDF model within that node to \circled{7} determine the \textbf{exact block number} that contains the key ($I_{IndexBlock}$) and \textbf{narrows down the search scope} for the last-mile search in the block using the LI model($I_{KV}$). \circled{8} Following this, \texttt{DobLIX} loads the block from storage in memory and \circled{9} executes the final search within the specified range in $KVs~Adrr$ stored in the blocks' metadata to find the exact offset of the target KV pair in the $KVs~Data$ (\S\ref{sec:design:lastmile}), and \circled{10} employs the retrieved address on the $KVs~Data$ to locate the actual KV. \circled{11} Finally, \texttt{DobLIX} measures the \textit{latency} of the current lookup query alongside the \textit{index size}, incorporating these measurements as feedback to refine the tuning agent (\S\ref{sec:design:agent}).

\subsection{Concept Overview}
\label{sec:design:overview}
The management of LSM data involves data partitioning and indexing phases (\S\ref{sec:rocksdb}), and as we established earlier, any optimization strategy, especially those involving LIs, should improve overall performance. As depicted in Fig.~\ref{fig:prev-designs}, block partitioning is intertwined with block indexing ($P_{Block} \not\perp I_{IndexBlock}$). Therefore, optimizing $I_{IndexBlock}$ requires the consideration of $P_{Block}$. In contrast, previous designs illustrated in Fig.~\hyperref[fig:prev-designs]{\ref*{fig:prev-designs}\{a,b,c\}} from earlier research have significant data access expenses due to the independence between the LI model and the data partitioning component.

\noindent As described in \S\ref{sec:li-storage}, within the traditional framework of LI modeling, $I(.)$ represents the result of approximating keys indexes drawn from an unknown distribution $\mathcal{D}_{keys}$ through practical optimization. Therefore, the classical design of the LI does not consider data access and the result coordinated by all data; however, in LSMs only a limited number of SST blocks reside in memory. In addition, the primary optimization objective is to minimize the error within the hypothesis spaces chosen, regardless of any secondary objectives. 

\texttt{DobLIX} aims to redefine LI models by integrating efficient block-based data access as a key objective. Enhances indexing performance by ensuring that the trained model accurately maps queries to the correct block, enabling the retrieval of only a single block while adhering to the optimal block size. A critical aspect of \texttt{DobLIX} is the relationship between the index approximation ($I_{IndexBlock}$) and block partitioning ($P_{Block}$), where a one-to-one correspondence is established between the index approximation and the segments within $P_{Block}$. This allows \texttt{DobLIX} to apply LI models that partition the key domain into segments, ensuring each segment corresponds to a specific block. The system performs a binary search on an array of offsets ($I_{IndexBlock}$) to find the start of each segment (i.e. block). Within each segment, it uses an index approximation ($I_{KV}$) to efficiently locate the keys. Since the trained index is based on the entire data in SST, the index model used for each block requires adjustment~(\S\ref{sec:design:lastmile}). This approach optimizes both block access and key retrieval, providing efficient indexing.

To achieve this, we introduce a dual-objective optimization approach for two distinct LI methodologies. The first method is based on PLA modeling~\cite{kipf2020radixspline}, while the second method employs PRA, based on the recursive model index~\cite{kraska2018case}. We delve into these methods in \S\ref{sec:design:pla} and \S\ref{sec:design:pra}.

\subsection{LI Approximation Methods}
\label{sec:design:li}
In this section, we present our LI algorithms that focus on dual-objective optimization to train the index model. Considering the importance of I/O performance in indexing, these algorithms are designed to partition the KV space based on their sizes and progressively systematically construct the index. Typically, we approximate the index for each segment using linear models. When it comes to searching for specific points during lookup processes, we employ a binary search on the points derived from the piecewise approximation to precisely pinpoint the required location, which is referred to as last-mile search. In the following, we elaborate on these algorithms in detail.

\begin{figure}[t]
  \centering
  \makebox{\includegraphics[width=0.9\columnwidth]{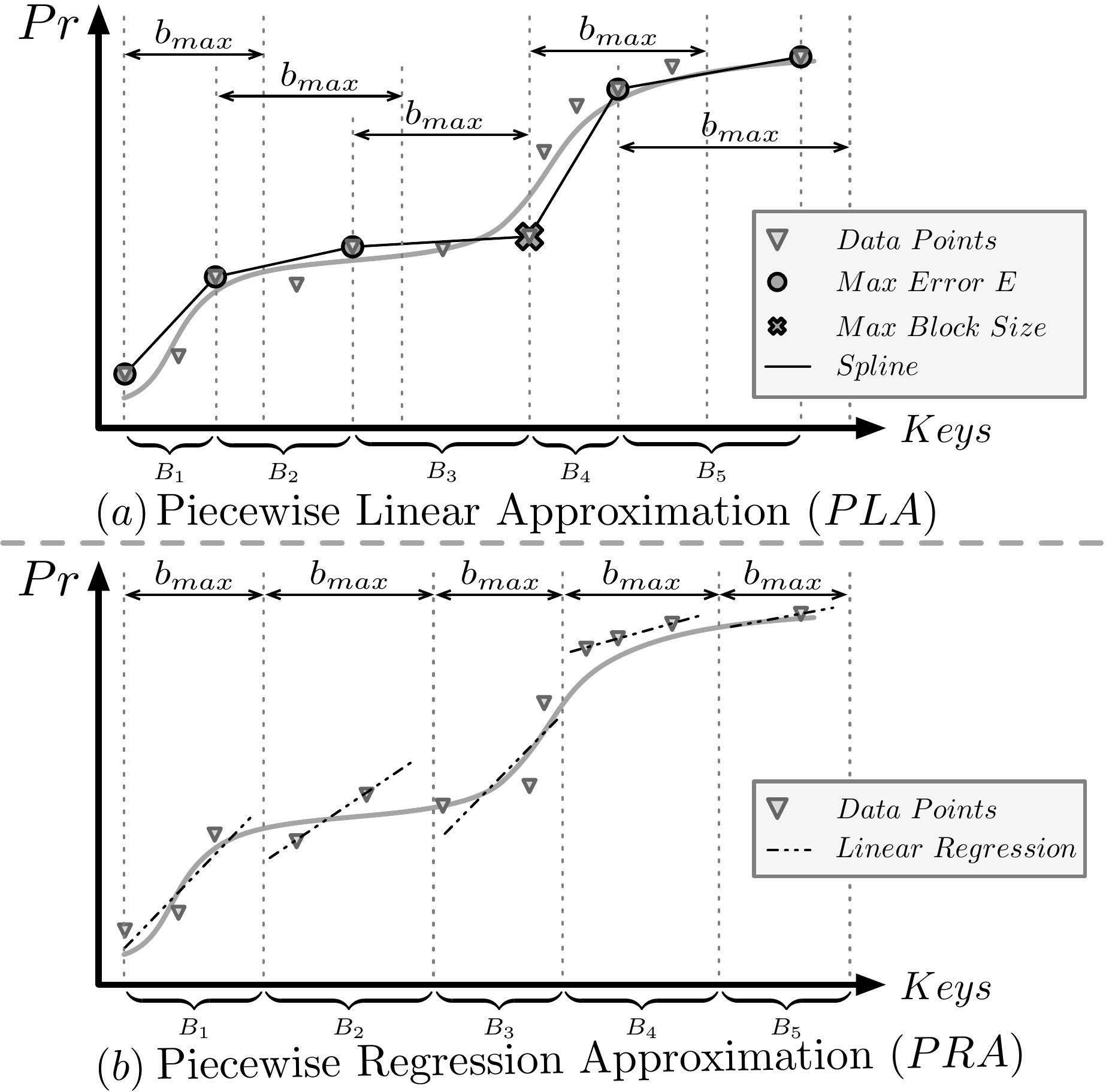}}

  \caption{\small{LI Models. $B_i$s represent the actual blocks added to SSTs.}}
  \label{fig:methods}
 
\end{figure}

\subsubsection{\textbf{Dual-Objective Optimization.}}
\label{sec:dual-objective-optimization}
Although various methods exist to tackle multi-objective optimization problems, we aim to prioritize the data access parameter more heavily than index lookup. Thus, we always finalize a block when its size exceeds the maximum block size $b_{max}$ by incorporating an additional pair of KVs. This ensures that the block sizes remain below $b_{max}$, even if the approximation error $E$ has not yet been achieved. In another words, our optimization to uncompromise the data access over the index lookup can be defined as: $\forall{B_i} \in P_{Block} : |B_i| \le b_{max}$. Note that both the configuration values $b_{max}$ and $E$ are given by the \textit{Tuning Agent} (\S\ref{sec:design:agent}).

\subsubsection{\textbf{Piecewise Linear Approximation (PLA).}}
\label{sec:design:pla}

In this method, the key space is divided into blocks $B_i$ using a linear approximation (spline), each block containing keys that share a common prefix. For each block $B_i$, a spline estimate of the positions is made, defined as $M_i(x) = a_i + m_i(x - x_i)$, where $a_i$ and $m_i$ are coefficients derived from spline points and $x_i$ is the initial key in block $B_i$. The binary search is then used around $M_i(k)$ to identify the precise index $I(k)$. For each block $B_i$, the next key is added to a new block ($B_{i+1}$) of the SST if (1) the approximation error $M_i(x)$ reaches the maximum threshold $E$, or (2) the condition $|B_i|\ge b_{max}$ indicates that including the new pair of KV would exceed the maximum block size allowed. This approximation process is illustrated in Fig.~\hyperref[fig:methods]{\ref*{fig:methods}a}. If the size of an added point exceeds $b_{max}$, a new block is created to maintain the optimal I/O (data access) performance of the previous block.
This mechanism results in a new set of spline points (the cross are spline points in Fig.~\hyperref[fig:methods]{\ref*{fig:methods}a}), introducing new blocks when the secondary optimization criterion is met, in addition to the standard blocks formed by reaching the maximum approximation error $E$. Alg.~\ref{alg:pla} showcases this approach, where $APE(line,set)$ calculates the maximum distance the points in $set$ can have from the given line $line$. In this context, $a_i$ is defined as $\mathcal{R}[i][0][1]$, $x_i$ corresponds to $\mathcal{R}[i][0][0]$, and $m_i$ is calculated as $\frac{\mathcal{R}[i+1][0][1] - \mathcal{R}[i][0][1]}{\mathcal{R}[i+1][0][0] - \mathcal{R}[i][0][0]}$, while the term $\text{offset}_i$ refers to $\mathcal{R}[i][1]$.

\begin{algorithm}[t]
\SetKwInput{KwResult}{Output}
\SetKwInput{KwIn}{Input}
\DontPrintSemicolon
\LinesNumbered
\SetAlgoNlRelativeSize{-1}
\caption{Dual-objective PLA}\label{alg:pla}
\footnotesize
\KwIn{Set of KVs $D$}
\KwResult{Radix Points $\mathcal{R}$}
$\mathcal{R}\gets [~], \quad index\gets 0, \quad \text{offset}\gets 0$ \;

$E,b_{max} \gets TuningAgent()$ \;

$B_{curr} \gets [(k_0,v_0)]$\;

\While{$(k,v) \in D$}{
    \If{$\vert B_{curr}\vert > b_{max}$}{
      $\mathcal{R} \gets \mathcal{R} + [B_{curr}.last]$\;
      
      $B_{curr} \gets [(k,index)]$\;
    }
  \eIf{$\vert B_{curr}\vert>1~\wedge~
  APE(\text{Line}(B_{curr}.\text{first},(k,index)), B_{curr})\ge~E$}{
      $\mathcal{R} \gets \mathcal{R}+[(B_{curr}.\text{last},\text{offset})]$ \;
      
      $\text{offset} \gets \text{offset} + |B_{curr}|$\;
      
      $B_{curr} \gets [(k,index)]$\;
  }{
  $B_{curr} \gets B_{curr} + [(k,index)]$\;
  }
  $index \gets index + 1$\;
}
$\mathcal{R} \gets \mathcal{R} + [(B_{curr}.\text{last},\text{offset})]$\;
\end{algorithm}

\subsubsection{\textbf{Piecewise Regression Approximation (PRA).}}
\label{sec:design:pra}

This approach involves initially dividing the key domain into segments of size $b_{max}$ and then approximating each segment linearly. The maximum approximation error for each segment, $E'$, is stored and utilized during the last-mile search phase (refer to Fig.~\hyperref[fig:methods]{\ref*{fig:methods}b}). Start scanning the KVs from the beginning; if incorporating the new KV into the existing segment exceeds $b_{max}$, start a new segment. Alg.~\ref{alg:par} performs this task in one pass, maintaining the integrity of each KV pair.

\noindent After partitioning the data ($P_{Block}$) using the optimal block size $b_{max}$, as outlined in Alg.~\ref{alg:pra}, a new model can be constructed for each partition through a linear approximation. The boundary points are retained for search tasks to determine the appropriate model for lookup queries, and $E'$ is stored in $\mathcal{E}$, respectively.

\begin{algorithm}[t]
\SetKwInput{KwResult}{Output}
\SetKwInput{KwIn}{Input}
\DontPrintSemicolon
\LinesNumbered
\SetAlgoNlRelativeSize{-1}
\caption{Partition~$P_b(.)$}\label{alg:par}
\footnotesize
\KwIn{Set of KVs $D$,  Maximum Block Size $b$}
\KwResult{Partition $\mathcal{P}$}

$t,\mathcal{P} \gets []~~\&~~ \text{offset}\gets 0$ \;

\While{$KV \in D$}{
    \If{$\vert t\vert + \vert KV\vert > b$}{
        $\mathcal{P} \gets \mathcal{P} + [(t,\text{offset})]$\;
        
        $\text{offset}\gets\text{offset} + |t| ~\&~ t \gets []$ \;
    }
    $t \gets t + [KV]$\;
}
$\mathcal{P} \gets \mathcal{P} + [(t,\text{offset})]$\;
\end{algorithm}

\begin{algorithm}
\SetKwInput{KwResult}{Output}
\SetKwInput{KwIn}{Input}
\DontPrintSemicolon
\LinesNumbered
\SetAlgoNlRelativeSize{-1}
\caption{Dual-objective PRA}\label{alg:pra}
\footnotesize
\KwIn{Set of KVs $D$}
\KwResult{Radix Points $\mathcal{R}$\newline Model Sets $\mathcal{M}$\newline Maximum Segment Errors $\mathcal{E}$}
$\mathcal{R},\mathcal{M}, \mathcal{E} \gets []$ \;

$b_{max} \gets TuningAgent()$ \;

$\mathcal{P} \gets P_{b_{max}}(D)$ \Comment{from Alg.~\ref{alg:par}}\;

\While{$Partition~par \in \mathcal{P}$}{
\If{$\vert par\vert>1$ }{
    $M\gets LinearRegression(par)$\;
    
    $\mathcal{M} \gets \mathcal{M} + [M]$\;

    $\mathcal{E} \gets \mathcal{E} + [APE(M, par)]$\;
}
    $\mathcal{R} \gets \mathcal{R} + [par.frist]$\;
}
$\mathcal{R} \gets \mathcal{R} + [par.last]$\;
\end{algorithm}

\begin{figure}[t]
  \centering
  \makebox{\includegraphics[width=0.9\columnwidth]{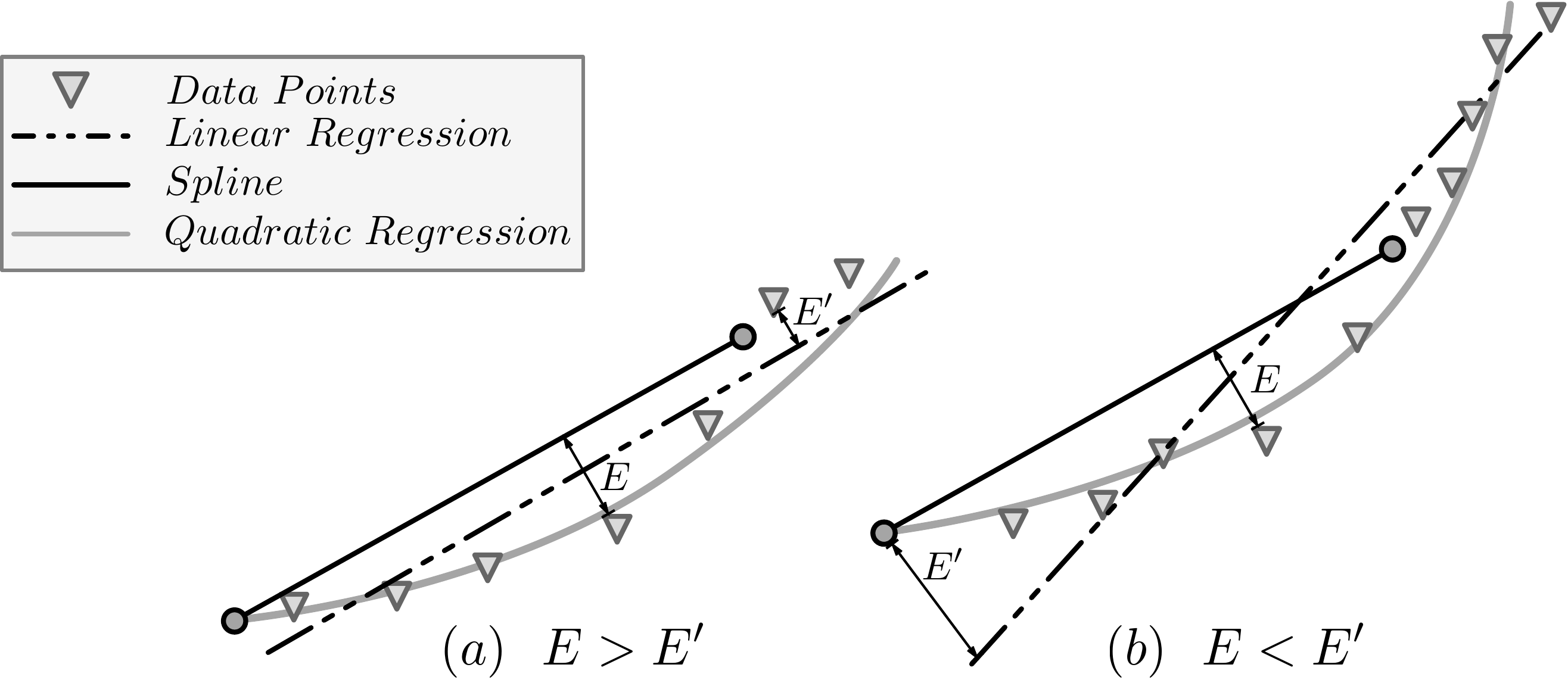}}
 
  \caption{\small{Comparison of PLA and PRA under different data distributions. (a) $E' < E$, indicating PRA performs better. (b) $E < E'$, indicating PLA performs better.}}
  \label{fig:compare-methods}
 
\end{figure}

\noindent\subsubsection{\textbf{Comparing PRA and PLA.}} In the context of building block approximations, both the PLA and the PRA rely on the scanning of data, resulting in a linear time complexity of $O(N)$, where $N$ is the number of data points. Each method utilizes closed-form formulas with a time complexity of $O(N)$ for computations within blocks: PLA determines maximum distances, $APE(.,.)$, to construct piecewise linear segments, while PRA computes two-dimensional regression, $LinearRegression(.)$, formulas within blocks.

\noindent With respect to space complexity, PLA requires one point per spline segment to be stored since the endpoint of one line serves as the beginning of the next. The distribution of data points influences PLA memory needs; for instance, with a uniform distribution, all data might fit within a single spline (provided its size is smaller than $b_{max}$), thus minimizing memory usage. In contrast, PRA must store both a point and a slope for each regression line, roughly doubling the memory requirement compared to PLA. The complexity of the number of regressions is $O(\frac{N}{b_{max}})$. Consequently, the memory comparison between PLA and PRA depends on the characteristics of the data: PLA may involve fewer blocks and needs just one point per block (linear segment), whereas PRA has to retain two parameters per block.

When comparing the behavior of PRA and PLA during lookups (\circled{7} in Fig.~\ref{fig:DobLIX-arch}), determining which method is superior is challenging, often leading to the interchangeable use of algorithms; in particular, we consider two scenarios that contrast PLA and PRA, as illustrated in Fig.~\ref{fig:compare-methods}, noting that both scenario~(a) and scenario~(b) can occur depending on the sizes of the KV pairs and the distribution of the keys. In the PRA model, the block is written in persistent storage once its size reaches the maximum block size $b_{max}$, whereas the PLA operates under two conditions for flushing: when the approximation error exceeds the error limit $E$, or when the block size reaches $b_{max}$ (the same maximum block size used in PRA). Consequently, depending on the arrangement and distribution of KV, the spline achievable with PLA can potentially lead to a maximum error $E$ that may be higher or lower than the maximum error $E'$ in PRA. As shown in Fig.~\hyperref[fig:compare-methods]{\ref*{fig:compare-methods}a}, PRA can result in $E > E'$, indicating a more accurate approximation than PLA. This directly affects the scope of the last-mile search and the overall efficiency of each algorithm.

This implies that the behavior of KVs beyond the block boundary determined by the maximum error $E$ plays a crucial role. If data points outside the block constrained by $E$ align with the regression trend of the points within the block, the approximation remains accurate. However, there might be situations where data points outside the block behave significantly differently from those within the block. In such cases, as clearly shown in Fig.~\hyperref[fig:compare-methods]{\ref*{fig:compare-methods}b}, this could lead to a less accurate linear regression in PRA, resulting in $E < E'$. This means that PRA has a lower performance than PLA in such scenarios. In some rare cases, all elements in $\mathcal{E}$~(Alg.~\ref{alg:pra}), denoted as $E'$ for each segment, are equivalent to $E$. In such scenarios, both algorithms exhibit identical last-mile search performance.

\subsection{Serialize and Deserialize Models}
\label{sec:design:serialization}
After training, the LI models are serialized and stored in a metadata block called the Index Block (see Fig.~\ref{fig:sst-format}). This process involves traversing the model tree structure via depth-first search (DFS), serializing each node byte-by-byte by writing the spline points data. During deserialization, the model reconstructs the tree by determining the type of each node (leaf or internal) and populating the corresponding data structure.

The learned indices generally outperform the binary search in terms of time complexity. Therefore, the size of the set of stored nodes for model approximation is smaller than $O(\log N)$ when reading, while writing the entire dataset takes $O(N)$. As a result, the serialized model size is asymptotically negligible (see \S\ref{sec:eval:storage}). In practice, RocksDB writes occur in the background flush and compaction process, and since the model is deserialized only once, overall read performance is significantly improved (see \S\ref{sec:eval:throughput}).

\subsection{Last-Mile Search Optimization}
\label{sec:design:lastmile}

\begin{figure}[t]
  \centering
  \begin{overpic}[width=\columnwidth]{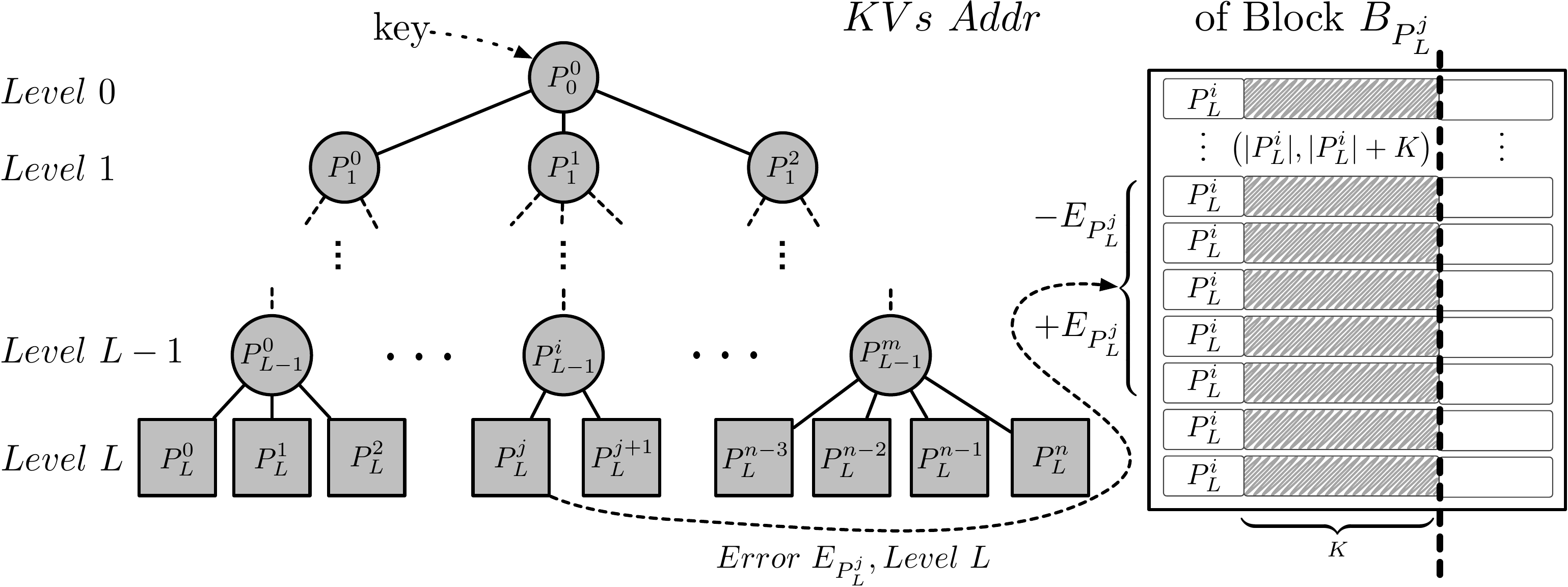}
    \put(66.9,35.5){\footnotesize (Fig.~\ref{fig:sst-format})}
  \end{overpic}
  \caption{\small{Last-Mile Search Optimization Flow}}
  \label{fig:last-mile-optimization}

\end{figure}

The final phase of the search process involves the last-mile search of the recovered block from storage (step \circled{9} in Fig.~\ref{fig:DobLIX-arch}). As illustrated in Fig.~\hyperref[fig:rocksdb-lookup]{\ref*{fig:rocksdb-lookup}c}, this step accounts for more than $40\%$ of the indexing latency in RocksDB. Fig.~\ref{fig:last-mile-optimization} shows how \texttt{DobLIX} optimizes its performance by restoring data from the LI model computation trajectory (step \circled{5} in Fig.~\ref{fig:DobLIX-arch}) to improve the last-mile search process. Initially, the target key is searched within the string LI structure, as explained in \S\ref{sec:li-storage}. The LI model in \texttt{DobLIX} subsequently provides: \textit{(1)} $Block~b_{P_L^j}$: The block containing the target KV pair.
\textit{(2)} $M(.)$: The LI estimation of the target KV pair index in $KVs~Addr$ (Fig.~\ref{fig:DobLIX-arch}).
\textit{(3)} $Level~L$: The level at which the target key was found in the LI model.
\textit{(4)} $Error~E_{P_L^j}$: The maximum range required to search for the target KV pair.

\noindent
{\small\textbf{Optimizing Search Range.}}
As described in \S\ref{sec:design:overview}, \texttt{DobLIX} necessitates a coordinate transformation to adapt the model output ($M(.)$) for indexing on the retrieved block ($B_{P_L^j}$). This is achieved by deducting the count of keys in the previous blocks (maintained as the parameter \say{offset} in the metadata of the block introduced in Algs.~\ref{alg:pla}\&~\ref{alg:par}):
{\small
\[
    I_{KV}(.)=M_{adj}(.) = M(.) - B_{P_L^j}.\text{offset}
\]
}
\noindent
To better illustrate the adjustment, we consider $SST_j$ in Fig.~\ref{fig:prev-designs}, in which the model is trained on the whole SST indexes. However, using the above adjustment, the coordination of model output for our solution (i.e., Fig.~\hyperref[fig:prev-designs]{\ref*{fig:prev-designs}d}) is transformed to the retrieved block $B_2$.

\noindent Subsequently, \texttt{DobLIX} performs a binary search within the specified model error range $E_{P_L^j}$ on $M_{adj}(.)$. This error range can be less than the maximum error $E$ if a spline in the PLA method reaches the maximum block size $b_{max}$. In such cases, \texttt{DobLIX} calculates the spline error and incorporates it into the model, transmitting this information to the last-mile search process to streamline the number of key comparisons.

\noindent
{\small\textbf{Optimizing String Comparison.}}
\texttt{DobLIX} further optimizes the comparisons by avoiding full key comparisons. \texttt{DobLIX} provides the node level where the key is found in its string-LI structure. This level in the tree indicates the common prefix string, $P_L^j$, in the key (Fig.~\ref{fig:last-mile-optimization}). Then \texttt{DobLIX} can ignore this common prefix, as all keys within the retrieved block share the same prefix. Additionally, \texttt{DobLIX} only compares string keys up to their $K$ bytes, ensuring that the key can be identified by comparing only the $K$ byte following the prefix within the error range. Given that $K$ generally consists of $8$ or $16$ bytes as explained in \S\ref{sec:li-on-strings} for shifts from demanding string comparisons to numerical comparisons. We perform an ablation study to analyze the impact of these optimization on \texttt{DobLIX} performance in \S\ref{sec:eval:ablation}.
\subsection{Tuning Agent}
\label{sec:design:agent}

\begin{figure}[t]
  \centering
\makebox{\includegraphics[width=0.95\columnwidth]{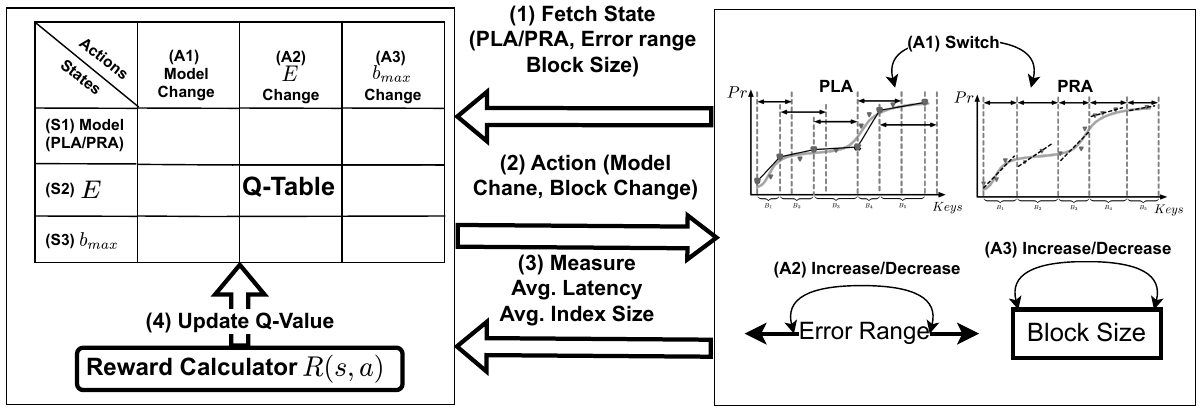}}

  \caption{\small{RL Tuning Agent Overview.}}
  \label{fig:rl-agent}

\end{figure}

\texttt{DobLIX} employs Q-learning~\cite{watkins1992q}, a lightweight Reinforcement Learning (RL) algorithm~\cite{mo2023learning}, to dynamically select model and fine-tune model and data access parameters. Fig.~\ref{fig:rl-agent} shows the overview of the RL tuning Agent. \texttt{DobLIX} determines three key parameters in the creation of SSTs and the training model: \textit{(1)} The choice between using the PLA vs. the PRA. \textit{(2)} The maximum error of the LI in the PLA method ($E$), and \textit{(3)} The maximum size of a block ($b_{max}$).

The state space for the Q-learning agent comprises the index model--Current Model which is $2$ states between PLA and PRA, Error values for PLA ranging from $32$ to $256$, doubling at each step, resulting in $4$ distinct values--and the block size--$4KB$ to $32KB$, doubling at each step, yielding $4$ distinct values, as recommended by RocksDB. The action space consists of: switching between PLA/PRA models, and incrementing, decrementing $b_{max}$, or $E$. We also forbid the actions to change $E$ when the model state is PRA as it does not consists of different $E$.


The reward function is defined below to consider both system latency and index storage:

\noindent
\scalebox{0.8}{%
\centering{
$R(s, a) = -\nu.Norm(AVG(latency)) - (1 - \nu).Norm(AVG(index~size)) $
}
}

\noindent This reward function is calculated as the negative sum of the normalized average end-to-end latency and the normalized average of SSTs index size. We use sigmoid normalization to ensure both latency and index size contribute to the reward on a comparable scale. The weighting parameter $\nu$ controls the relative importance of latency and index size in determining the overall reward (evaluated in \S\ref{sec:eval:param}).

\begin{algorithm}[t]
\DontPrintSemicolon  
\caption{System Tuning Agent}\label{alg:tune}
\footnotesize
$Q \gets \text{initializeQTable()}$ \;
$\alpha, \gamma, \epsilon, \nu \gets \text{initializeVariables()}$ \;
\While{}{
    Fetch state $s_{t-1}$ \;
    $L_{t}, I_{t} \gets \text{fetchAverageLatencyAndSSTIndexSize()}$ \;
    $R_{t} \gets \text{calculateReward}(L_{t}, I_{t})$ \;
    Observe state $s_{t}$ \;
    $a' \gets \arg\max_{a \in \text{Actions}} Q(s_{t},a)$ \;
    $Q(s_{t-1}, a_{t-1}) \gets (1-\alpha)Q(s_{t-1}, a_{t-1}) + \alpha\big(R_{t} + \gamma Q(s_{t},a')\big)$ \;
    $A_t \gets \text{getAvailableActions}(s_t)$ \;
    \eIf{$\text{generateRandNumber()} < \epsilon$}{
        $a_t \gets \text{getRandomAction}(A_t)$ \;
    }{
        $a_t \gets \arg\max_{a \in A_t} Q(s_t,a)$ \;
    }
    tuneSystem($a_t$) \;
    $s_t \gets s_{t+1}$ \;
    $\epsilon \gets \text{updateEpsilon}(\epsilon)$ \;
}

\end{algorithm}

Alg.~\ref{alg:tune} outlines the agent tuning and execution process. The learning procedure begins by configuring the RL hyperparameters: $\alpha$ denotes the learning rate, while $\gamma$ signifies the discount factor that balances immediate and future rewards. An epsilon decay strategy is used by initially setting a high exploration rate ($\epsilon$) to investigate various actions, gradually lowering this value in successive iterations to exploit the optimal actions.
At every $20$ SSTs creation, the following steps occur: First, the agent obtains the average latency of the reads that have reached this level, as well as the index size of the previously created SST. Then, it measures the reward and updates it for the chosen action in the previous state. The agent then gets the available actions at that state. Then, depending on the value of $\epsilon$, the agent explores a new action or chooses the best action from the Q-table. In addition, we implement a reset mechanism for the RL agent that reverts $\epsilon$ to its initial value if the distribution of incoming KVs undergoes a significant change. This ensures optimal exploration of the space under the new conditions.

\section{EVALUATIONS}
\label{sec:eval}
This section presents the results of the \texttt{DobLIX} evaluation, emphasizing its advantages over current leading methods. We design the evaluations to answer the following questions: 1) What performance advantages does \texttt{DobLIX} offer? (\S\ref{sec:eval:perf}) 2) How does the length of key-value pairs affect \texttt{DobLIX} and other baseline systems? (\S\ref{sec:eval:kvlen}) 3) What is the Storage and Write amplification of \texttt{DobLIX}? (\S\ref{sec:eval:storage}) 4) How does the RL agent adjust its underlying parameters? (\S\ref{sec:eval:param})

\subsection{Experimental Setup}
\label{sec:experimental_setup}

\noindent\subsubsection{\textbf{Environment.}}
We implemented DobLIX in C++17 with RocksDB 8.1.1, compiled using GCC 9.4.0. We ran evaluations on Ubuntu 20.04 with a 64 core AMD Ryzen Pro 5995WX (2.7GHz) and 256GB DDR4 RAM. The storage device used was a SAMSUNG 980 Pro 2TB M.2 NVMe SSD. This NVMe SSD offers fundamental read and write performance measures as follows: sequential read speeds of 7,000 MB/s, random read speeds of 1,237K IOPS, sequential write speeds of 5,000 MB/s, and random write speeds of 172.5K IOPS. A single experiment was performed on a Seagate BarraCuda 4TB SATA SSD. We allocate 4 background threads to RocksDB and set the \texttt{\small max\_background\_compactions} and \texttt{\small max\_background\_flushes} options to 4, as prior studies~\cite{TridentKV2022,Bourbon2020,sarkar2023lsm} and RocksDB best practices suggest this as a balanced trade-off between compaction efficiency and I/O contention, ensuring stable read performance. Each compaction and flush task runs in a separate thread, as RocksDB dynamically assigns work within the allocated thread pools rather than using a fixed number of threads per worker.

\noindent\subsubsection{\textbf{Datasets.}} 
We utilize four real-world datasets sourced from the Search on Sorted Data Benchmark (SOSD)\cite{kipf2019sosd} along with two synthetic datasets for evaluating \texttt{DobLIX}. These datasets from prior studies~\cite{TridentKV2022,Bourbon2020,ding2020alex} each hold 64M key-value pairs. The size of the key and the value are set according to the experiments used in our baselines (\cite{Bourbon2020,TridentKV2022}) as $8$ byte and $64$ byte. We also conduct experiments with other fixed key and value sizes, and also with varying key and value sizes to reflect real-world applications.
The following sections provide specific information about each dataset. \textit{WIKI~\cite{WikiTS}.} Edit timestamps for Wikipedia articles. \textit{AMZN}: Popularity of book sales collected from Amazon. \textit{FB~\cite{FB}}: An upsampled version of a Facebook user ID dataset.  \textit{OSM~\cite{OSM}}: Uniformly sampled locations as Google CellIds.  \textit{LOGN}: This synthetic dataset is generated from a lognormal distribution with parameters $\mu=0$ and $\sigma=2$, multiplied by $10^9$ and rounded down to the nearest integer. UNI: Synthetic data sampled uniformly between $0$ and $10^{16}$.

\noindent
\subsubsection{\textbf{Workloads.}}
We evaluate \texttt{DobLIX} with four different workloads, each consisting of $10$ million operations. \textit{Read-Only~(\textbf{RO}):} Focuses solely on read operations. \textit{Read-Heavy~(\textbf{RH}):} Emphasizes reads ($90\%$) with a smaller proportion of inserts ($10\%$). \textit{Balanced~(\textbf{BA}):} Involves an equal split between reads and inserts ($50\%$ each). 
Write-Heavy~(\textbf{WH}): Emphasizes insert operations ($90\%$) over read operations ($10\%$).
We also use YCSB~\cite{cooper2010benchmarking} real benchmarks to evaluate \texttt{DobLIX}. In all workloads, the search key is selected randomly from the existing set of keys in the index with a Zipfian distribution~\cite{cormode2008finding}, unless a different request distribution is specified.

\noindent
\subsubsection{\textbf{Baselines.}} 
\label{sec:eval:baselines}
We use these schemes as comparison baselines. \textbf{\textit{{RocksDB}}}~\cite{rocksdbpaper}: An embedded high-performance KV-store used as the storage engine. We utilize the default indexing mechanism of RocksDB.
\textbf{\textit{{Bourbon}}}~\cite{Bourbon2020}: A KV store with an LI that accelerates lookups by understanding the distribution of keys. Based on WiscKey~\cite{lu2017wisckey}, a LevelDB~\cite{leveldb} variant, Bourbon retains fixed block sizes, but might need to load several blocks (See Fig.~\hyperref[fig:prev-designs]{\ref*{fig:prev-designs}A}). The Bourbon source code is publicly accessible~\cite{BourbonGithub}. We incorporate Bourbon indexing method into our RocksDB configuration. \textbf{\textit{{TridentKV}}}~\cite{TridentKV2022}: an LI variant of RocksDB that aims to retrieve KV pairs by loading only one data block. TridentKV modifies the size of the data blocks, potentially significantly increasing the size of the block, which can affect the lookup performance when loading a large block (see Fig.~\hyperref[fig:prev-designs]{\ref*{fig:prev-designs}C}). TridentKV code is available as open-source~\cite{TridentKVGithub} and is built on top of RocksDB. We adopt their implementation in our evaluations.

\begin{figure*}
  \centering
  \makebox[\textwidth][c]{\includegraphics[width=0.99\textwidth]{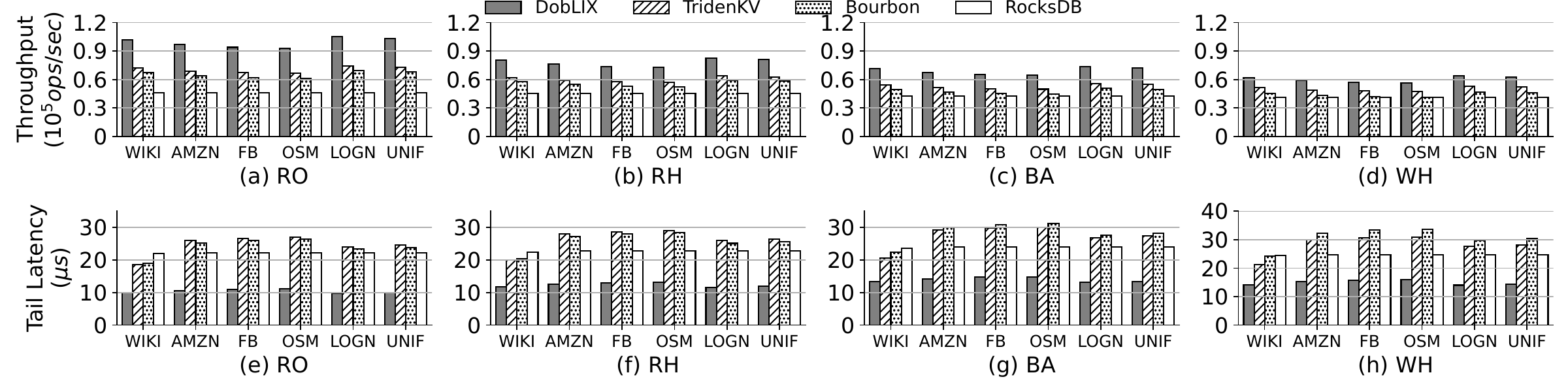}}
  \caption{\small{Throughput Comparison (upper Figure) and Tail Latency Comparison (lower Figure).}}
  \label{fig:eval-throughput}
  \label{fig:eval-latency}
\end{figure*}

\noindent
\subsubsection{\textbf{Metrics.}} We use the following metrics to evaluate \texttt{DobLIX} and all other baselines. \textbf{\textit{{Throughput:}}} Average rate of operations per second. \textbf{\textit{ {Latency:}}} Average end-to-end latency across all operations, excluding the slowest $1\%$ ($99^{th}$ percentile). \textbf{\textit{ {Tail Latency:}}}  The average latency of the $5\%$ slowest operations. \textbf{\textit{ {Index size:}}} LI and index block size. \textbf{\textit{{Compaction time:}}} The average duration taken to create SSTs in the compaction process.

\noindent
\subsubsection{\textbf{Parameters.}}
\label{sec:parameters}
By default, all methods adhere to the default configuration settings of RocksDB. The default configurations of Bourbon and TridentKV are also employed.
For certain parameters such as the chosen learning method (either PLA or PRA), the maximum error bound, and the maximum block size, the tuning agent is responsible for making decisions. Initially, we use a $1\%$ sample from the datasets to train this agent, after which we incorporate the agent into the system. In the case of the PLA method, we use the same setup as RSS~\cite{rss} to configure a dynamic radix table with $18$ bits for the first level, $12$ bits for the second level and $8$ bits for the third level and beyond. 

Regarding the hyperparameters of the RL agent $\alpha$ and $\gamma$ (see \S\ref{sec:design:agent}), we perform a sensitivity test with values $0.2$, $0.5$ and $0.8$. We observe that optimal results are achieved by assigning a low value of $0.2$ to $\alpha$ and a high value of $0.8$ to $\gamma$. We also set the initial value of $\epsilon$ at $0.99$, with its minimum value being $0.02$, allowing for a low exploration rate even after the training period (\S\ref{sec:exp:rl-agent-tuning}). We set $\nu$ in the reward function as 1 to optimize solely on performance.

\subsection{Performance}
\label{sec:eval:perf}
In this section, we detail a set of experiments designed to evaluate \texttt{DobLIX} performance, alongside a comparison with baseline systems.
The initial experiment (\S\ref{sec:eval:throughput}) highlights that \texttt{DobLIX} improves throughput by up to $1.41\times$, $1.52\times$, and $2.21\times$ relative to TridentKV, Bourbon, and RocksDB, respectively.
In the second experiment (\S\ref{sec:eval:tail}), \texttt{DobLIX} achieves up to $2.67\times$ better throughput in terms of tail latency.
The third experiment (\S\ref{sec:eval:breakdown}) examines the optimization of \texttt{DobLIX} lookup latency components.
The fourth experiment (\S\ref{sec:eval:ycsb}) showcases \texttt{DobLIX} enhanced performance on the realistic YCSB benchmarks.
Finally, the last experiment (\S\ref{sec:eval:req-distr}) shows \texttt{DobLIX} performance across varying request distributions. 

\subsubsection{\textbf{Throughput.}}
\label{sec:eval:throughput} 
This section displays the throughput when subjected to a specific workload on different datasets.

\noindent
{\small\textbf{RO Workload}} illustrated in Fig.~\hyperref[fig:eval-throughput]{\ref*{fig:eval-throughput}a}, where \texttt{DobLIX} demonstrates the highest throughput, achieving a maximum of $105K~ops/sec$ using the LOGN dataset. Compared to TridentKV, Bourbon, and RocksDB, \texttt{DobLIX} increases the average throughput by $1.41\times$, $1.57\times$, and $2.21\times$, respectively. {\small\textbf{RH Workload}} throughput depicted in Fig.~\hyperref[fig:eval-throughput]{\ref*{fig:eval-throughput}b}, where \texttt{DobLIX} consistently achieves the highest throughput across all datasets, peaking at $82K~ops/sec$. Compared to TridentKV, Bourbon, and RocksDB, \texttt{DobLIX} increases the average throughput by $1.27\times$, $1.36\times$, and $1.22\times$, respectively. Despite the marginal reduction in throughput from write operations due to the slight training overhead for new SSTs, \texttt{DobLIX} and other LIs still maintain superior throughput compared to RocksDB indexing approach. {\small\textbf{BA Workload}} in Fig.~\hyperref[fig:eval-throughput]{\ref*{fig:eval-throughput}c} illustrates that \texttt{DobLIX} consistently surpasses other systems in throughput, with an average enhancement of $1.29\times$, $1.27\times$, and $1.31\times$ over TridentKV, Bourbon, and RocksDB, respectively.
{\small\textbf{WH Workload}} in Fig.~\hyperref[fig:eval-throughput]{\ref*{fig:eval-throughput}d} shows that \texttt{DobLIX} achieves the highest throughput, reaching $53K~ops/sec$. When compared to TridentKV, Bourbon, and RocksDB, \texttt{DobLIX} enhances the average throughput by $1.16\times$, $1.14\times$, and $1.04\times$, respectively. \texttt{DobLIX}'s slight advantage over RocksDB is explained in \S\ref{sec:design:serialization}.

\noindent
\textit{\small\textbf{Takeaway}:}
\texttt{DobLIX} enhances read throughput through co-optimized data access and index lookups. Compared to Bourbon (that loads multiple blocks per lookup) and TridentKV (features excessively large block sizes), \texttt{DobLIX} minimizes read amplification by ensuring single-block reads with optimal block sizes (less than $b_{max}$) per query, while maintaining comparable write amplification to RocksDB. \texttt{DobLIX} also achieves write superior to Bourbon and TridentKV due to its smaller index size (see \S\ref{sec:eval:storage}).


\subsubsection{\textbf{Tail Latency}} 
\label{sec:eval:tail}
Analyzing tail latencies of different methods is vital because it defines the performance expectations for users and applications regarding their back-end key-value storage engine under the most challenging scenarios.

\noindent
{\small\textbf{RO Workload.}}
Fig.~\hyperref[fig:eval-latency]{\ref*{fig:eval-latency}e} shows the tail latency results for RO workloads, indicating that \texttt{DobLIX} exhibits the lowest tail latency. Notably, \texttt{DobLIX} improvement in tail latency surpasses the improvements observed in throughput. Compared to TridentKV, Bourbon, and RocksDB, \texttt{DobLIX} achieves improvements of $1.85\times$, $1.89\times$, and $2.13\times$, respectively. This difference is due to the high read amplification and index lookup inefficiency of other methods, explained in \S\ref{sec:eval:perf}.

\noindent
{\small\textbf{Workloads with Write.}} Fig.~\hyperref[fig:eval-latency]{\ref*{fig:eval-latency}\{f,g,h\}} present the tail latency results for RH, BA, and WH workloads. These figures illustrate that \texttt{DobLIX} achieves the lowest tail latency across all workload types. In terms of tail latency, RocksDB indexing outperforms both Bourbon and TridentKV, which employ LIing without optimizing for the data access phase. TridentKV exhibits the highest tail latency among these workloads, showing up to $2.67\times$ greater tail latency than \texttt{DobLIX} due to its large blocks, which negatively affect both read and write operations. Bourbon also experiences elevated tail latency in write-intensive workloads due to its dependence on a garbage collection mechanism\cite{Bourbon2020}, which degrades its performance during this period.

\noindent\textit{\small\textbf{Takeaway}:} The result shows that \texttt{DobLIX} has lower tail latency while competing with its opponents in the ordinary time span. 

\begin{figure}[t]
  \centering
  \makebox[\columnwidth][c]{\includegraphics[width=\columnwidth]{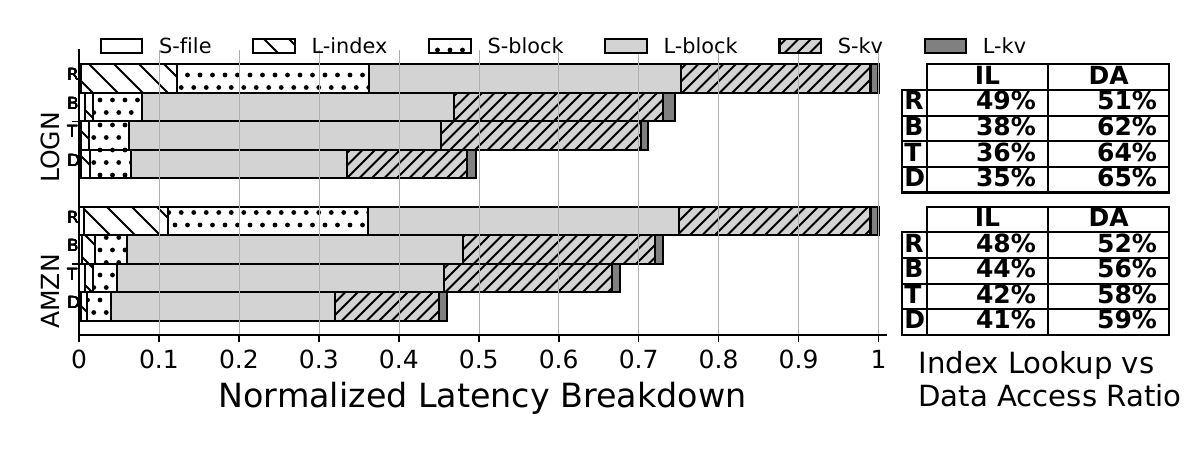}}
  \caption{\small{Latency Breakdown (Normalized) of \texttt{DobLIX} (`D'), TridentKV(`T'), Bourbon(`B'), and RocksDB (`R'). `L' and `S', on the legend, stand for load and seek, respectively. Tables show the relative time in Index lookup (`IL') vs. Data acess (`DA') for all methods.}}
  \label{fig:eval-breakup}
\end{figure}

\subsubsection{\textbf{Lookup Query Latency Breakdown}}
\label{sec:eval:breakdown}
In Fig.~\ref{fig:eval-breakup}, the breakdown of the latencies of systems in RH workloads is presented using the AMZN and LOGN datasets. Latencies are normalized by RocksDB latency.
\texttt{DobLIX} achieves average latency improvements of up to $23\%$, $27\%$, and $56\%$ compared to TridentKV, Bourbon, and RocksDB. Certain components, such as the Seek file or $I_{SST}$ (S-file) and Load KV (L-KV), show minimal changes, as these steps are consistent across all methods. The load index block (L-index) sees an improvement of $10.35\%$ on average compared to RocksDB, attributed to \texttt{DobLIX} LI methods. \texttt{DobLIX} also improves the lookup of blocks or $I_{IndexBlock}$ (S-block) by up to $22\%$ compared to RocksDB. Significant enhancements compared to Bourbon and TridentKV are seen in block loading (L-block) and locating the target KV within the block or $I_{KV}$ (S-KV). \texttt{DobLIX} multi-objective optimization strategy targets block size and precise block retrieval, resulting in a decrease in average block loading time by as much as $12.2\%$ compared to other methods. In terms of the KV last-mile search (S-KV), \texttt{DobLIX} demonstrates an average reduction of 11.1\% due to two optimizations: the elimination of key prefixes and the narrowing of the error bound range (as discussed in \S\ref{sec:design:lastmile}).

\noindent
\textbf{Index Lookup vs. Data Access.}
Tables on Fig.~\ref{fig:eval-breakup} presents the relative time breakdown between index lookup and data access across different methods. The results demonstrate that LI methods require significantly less time for indexing compared to RocksDB's native implementation, with improvements reaching up to 9\% for the LOGN dataset. Also, \texttt{DobLIX} improvement over LI makes it achieve the lowest index lookup time among LI methods.

\begin{figure}[b]
  \centering
  \makebox[\columnwidth][c]{\includegraphics[width=\columnwidth]{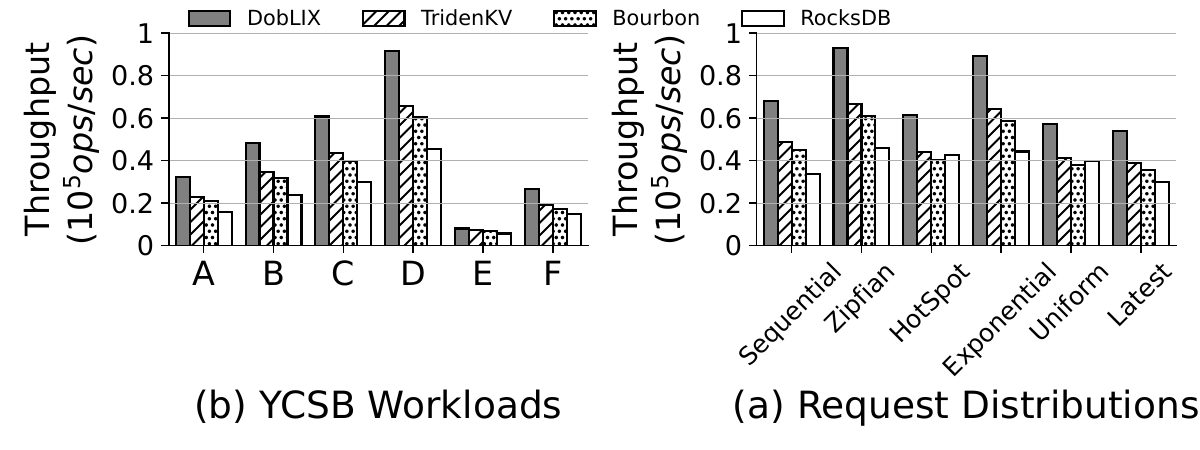}}
  \caption{\small Throughput Comparison on YCSB macrobenchmarks and various distribution workloads.}
  \label{fig:ycsb-distr}
\end{figure}

\subsubsection{\textbf{YCSB Macrobenchmarks}}
\label{sec:eval:ycsb}
Fig.~\hyperref[fig:ycsb-distr]{\ref*{fig:ycsb-distr}a} illustrates the throughput performance of \texttt{DobLIX} and the three baseline systems on various YCSB workloads. In particular, \texttt{DobLIX} consistently achieves the highest throughput on the six workloads.
In the read-heavy YCSB\{B,C,D\} workloads, \texttt{DobLIX} demonstrates superior throughput of $481K$, $607K$, and $916K~ops/sec$, respectively. On average, \texttt{DobLIX} increases throughput by $1.32\times$, $1.42\times$, and $2.02\times$ compared to TridentKV, Bourbon, and RocksDB, respectively.
For balanced workloads YCSB\{A,F\}, \texttt{DobLIX} reaches throughput levels of $320K$ and $264K$ ops/sec, respectively, outperforming other techniques with an average increase in throughput of $56\%$.
In the YCSB-E scan-focused benchmark, \texttt{DobLIX} achieves the highest throughput of $80K$ operations per second. However, its gains over TridentKV, Bourbon, and RocksDB are relatively modest at $9.2\%$, $16.7\%$, and $42.8\%$, respectively. This relative performance is because the LIs enhance point queries rather than scan operations. During the scan phase of the range queries, all techniques exhibit uniform performance.

\subsubsection{\textbf{Various Request Distributions}}
\label{sec:eval:req-distr}
Fig.~\hyperref[fig:ycsb-distr]{\ref*{fig:ycsb-distr}b} shows the throughput of \texttt{DobLIX} and the three baselines in RO workloads in the OSM dataset on six different request distributions. The figure shows that \texttt{DobLIX} maintains its superior read performance in the six request distributions. On average, \texttt{DobLIX} achieves $39.4\%$, $52.1\%$, and $78.9\%$ higher throughput compared to TridentKV, Bourbon, and RocksDB.

\subsection{Impact of Key and Value Length}
\label{sec:eval:kvlen}
In \S\ref{sec:eval:perf}, we evaluated \texttt{DobLIX} performance with fixed $8$-byte keys and $64$-byte values. Here, we first show \texttt{DobLIX} performance by changing the key and value sizes while maintaining their constancy during the experiments. Then, we perform more realistic experiments in which the key and value sizes vary within each experiment (details explained in \S\ref{sec:var-kv-lsm}).

\subsubsection{\textbf{Fixed-sized key-values}}
\begin{figure}[t]
  \centering
  \makebox[\columnwidth][c]{\includegraphics[width=\columnwidth]{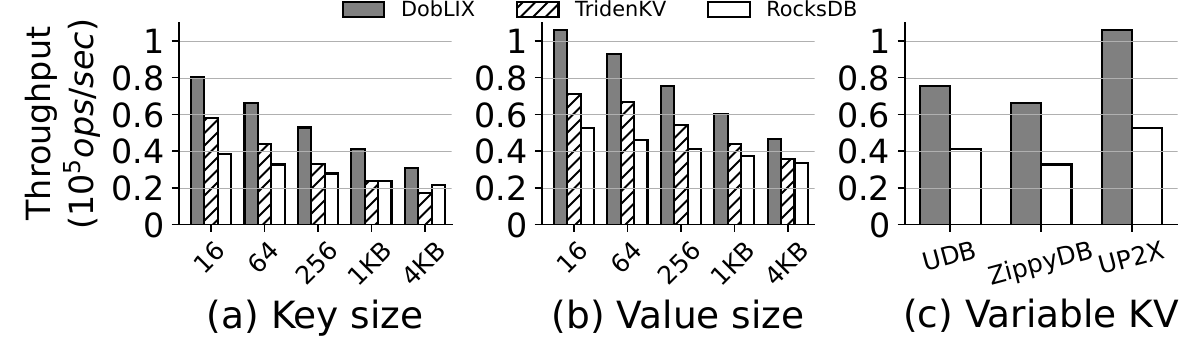}}
  \caption{\small{Impact of various key-value sizes.}}
  \label{fig:eval-kv-size}
 
\end{figure}

Fig.~\hyperref[fig:eval-kv-size]{\ref*{fig:eval-kv-size}a} shows how each method is affected by key size. Bourbon was excluded due to its 8-byte key limit. \texttt{DobLIX} delivers superior throughput on the OSM read-only workload by efficiently managing key prefixes, regardless of key size growth. However, TridentKV throughput drops drastically due to creating large block sizes by increasing the key size. The performance gap between \texttt{DobLIX} and RocksDB narrows with increasing size, as larger key-value loads impact both methods.
Fig.~\hyperref[fig:eval-kv-size]{\ref*{fig:eval-kv-size}b} shows the impact of various fixed value sizes on throughput. The results show that \texttt{DobLIX} has the highest throughput up to a value of $1KB$. For the value size $4KB$, Bourbon works better than TridentKV as it disaggregates KVs, which is efficient for large value sizes.

\subsubsection{\textbf{Variable-sized key-values}}
In this section, we conduct practical experiments utilizing variable KV sizes. We used the mean and standard deviations of the KVs of three RocksDB applications (Table~\ref{tab:kv-avg-sd}). As none of the previously established index methods can manage variable KVs, our comparison is exclusively with RocksDB. Fig.~\hyperref[fig:eval-kv-size]{\ref*{fig:eval-kv-size}c} shows the experimental results for these KV sizes. It illustrates that \texttt{DobLIX} functions effectively with variable key-value sizes and achieves $1.32\times$, $1.38\times$, and $1.62\times$ in key-value sizes such as UDB, ZippyDB, and UP2X.

\subsection{Storage and Write Amplification}
\label{sec:eval:storage}
\label{sec:storage}
\begin{figure}[t]
  \centering
  \makebox[\columnwidth][c]{\includegraphics[width=\columnwidth]{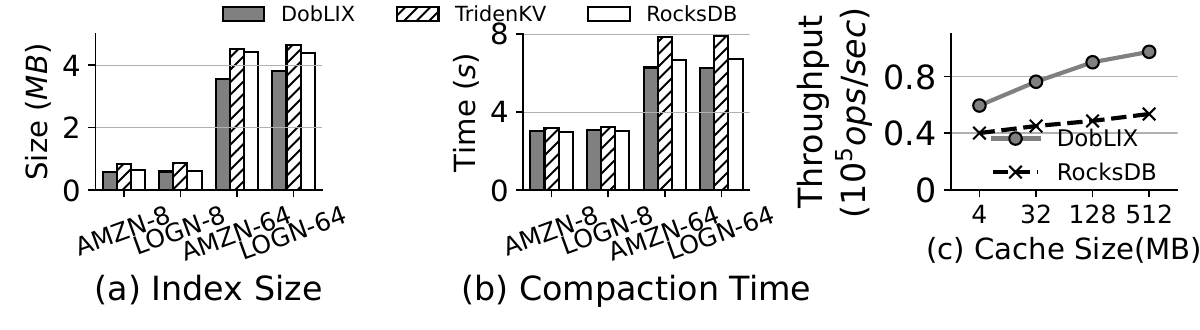}}
  \caption{\small{(a) Average index size, (b) Compaction time, (c) Effect of cache size on throughput.}}
  \label{fig:eval-mem-construction}
\end{figure}

Fig.~\hyperref[fig:eval-mem-construction]{\ref*{fig:eval-mem-construction}a} shows the index size for the AMZN and LOGN datasets with $8$- and $64$-byte key sizes. We omit Bourbon as it only works with $8$-byte numerical keys. The index block for each method will be written to storage beside the actual key values, hence index size has a direct relation with storage (the total space required on persistent media) and write amplification (the ratio of actual writes to storage versus logical writes requested by the system). The figure shows that in $8$-byte keys, the index size of \texttt{DobLIX} and RocksDB has a negligible difference, while both are less than TridentKV. However, for $64$-byte keys, \texttt{DobLIX} improves indexing by $25.9\%$ on average. This is because \texttt{DobLIX} removes the longest common prefix from each cluster of keys on each node (see \S\ref{sec:design:lastmile}), and each node only keeps $8$-byte numerical data. This shows that \texttt{DobLIX} has the lowest index size, and since it does not add any other data except the serialized LI model, it even decreases the storage and write amplification compared to RocksDB native indexing.

\subsection{Compaction Time}
SSTs in \texttt{DobLIX} and other baselines are built during background flush and compaction.~\cite{yu2024caas}. Fig.~\hyperref[fig:eval-mem-construction]{\ref*{fig:eval-mem-construction}b} demonstrates the average compaction time to build SSTs on the four different workloads and datasets mentioned in \S\ref{sec:storage}.
The data indicate that the \texttt{DobLIX} LI model does not add overhead during the SST construction compared to RocksDB. Notably, with $64$-byte keys, \texttt{DobLIX} construction time is $5.9\%$ shorter than RocksDB, due to the smaller \texttt{DobLIX} index size (Fig.~\hyperref[fig:eval-mem-construction]{\ref*{fig:eval-mem-construction}a}). Additionally, compared to TridentKV, \texttt{DobLIX} reduces construction time by $24.7\%$, attributed to TridentKV lack of prefix removal optimization in its string model. Specifically, \texttt{DobLIX} average SST construction time for $64$-byte keys is $6.2$ seconds, with model training taking $302$ ms ($4.8\%$ of the total time) and serialization and metadata writing taking $237$ ms ($3.7\%$ of the total time). In contrast, TridentKV model training time is $817$ ms ($10.4\%$ of the total time), and serialization plus metadata writing time is $414$ms ($5.1\%$ of the total time). 

\subsection{Effect of Cache Size}
Fig.~\ref{fig:eval-mem-construction}c demonstrates how cache size affects system throughput using the RO workload with Zipfian distribution on the AMZN dataset. While RocksDB's default configuration employs a 32MB cache, both \texttt{DobLIX} and RocksDB benefit from larger caches through reduced disk accesses via prefetching of data and index blocks. \texttt{DobLIX} achieves significant gains, outperforming RocksDB by $81\%$ at 512MB cache size. This substantial improvement occurs because DobLIX's efficient index lookup becomes more pronounced when caching reduces data access overhead, aligning with observations by Lu~et~al.~\cite{TridentKV2022}.

\subsection{Parameter Tuning}
\label{sec:eval:param}
\label{sec:exp:rl-agent-tuning}

\begin{figure}[t]
  \centering
  \makebox[\columnwidth][c]{\includegraphics[width=0.9\columnwidth]{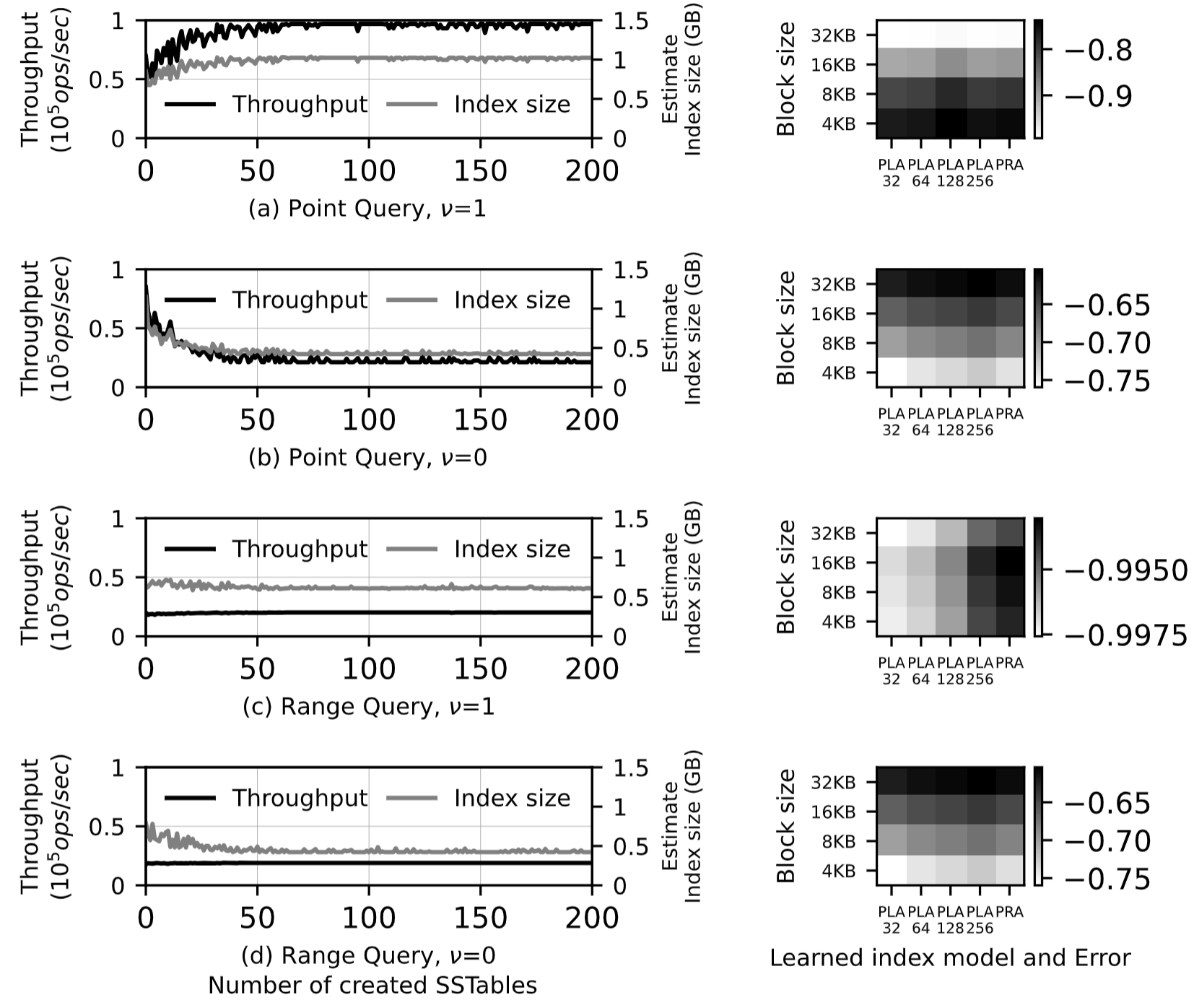}}
  \caption{\small{RL agent parameter tuning. Each row shows two plots for one specific workload and reward parameter, The right plot shows the improvement in the throughput and estimated index size during the workload. The left plot shows the reward heatmap at the end of the experiment. Darker colors have higher rewards.}}
  \label{fig:eval-rl}
\end{figure}

In this section, we present the results of the RL agent for parameter tuning and analyze the impact of parameters on throughput and total index size. The evaluation uses the AMZN dataset under two RH workloads: (1) point-query with a Zipfian distribution, and (2) range queries over $100$ consecutive KV pairs with a uniform distribution. We also vary the $\nu$ parameter in the RL reward function (\S\ref{sec:design:agent}) between $0$ and $1$ to evaluate trade-offs between index storage footprint and throughput optimization. The left side of Fig.~\hyperref[fig:eval-rl]{\ref*{fig:eval-rl}} illustrates throughput between two RL episodes, and the estimated total index size after each episode. The index size is estimated using a linear approximation based on current keys added to a total of $64$ million keys in the dataset. The right side of the figure displays the RL agent Q-table rewards, highlighting the best parameter configurations for each workload.

\noindent
\textbf{Throughput Optimization.} The throughput of point queries (Fig.~\hyperref[fig:eval-rl]{\ref*{fig:eval-rl}a}) increases from $45K$ to $93K$ ops/sec after $50$ episodes. This improvement is achieved by tuning parameters to prioritize throughput at the cost of increasing storage size. The heatmap shows that smaller block sizes yield higher rewards because they reduce read amplification, minimizing the size of the retrieved block per query. Among the index models, PLA with a maximum error of $128$ achieves the best performance. A smaller error increases the block count, while a larger error (i.e., $256$) raises the cost of intra-block last-mile searches.
For range queries (Fig.~\hyperref[fig:eval-rl]{\ref*{fig:eval-rl}c}), parameter rewards are less sensitive compared to point queries since range query performance is dominated by sequential KV iteration. However, a block size of $16KB$ paired with the PRA model achieves the highest reward. This size strikes a balance, being more efficient than $8KB$ since the data being read is more than $8KB$, so reading $16KB$ blocks lowers the amount of metadata reads. Additionally, the PRA model outperforms PLA with different error ranges at the $16KB$ block size, as this specific setting has a characteristic similar to Fig.~\hyperref[fig:compare-methods]{\ref*{fig:compare-methods}b} where $E' < E$.

\noindent
\textbf{Storage Optimization.} Fig.\hyperref[fig:eval-rl]{\ref*{fig:eval-rl}b} and Fig.\hyperref[fig:eval-rl]{\ref*{fig:eval-rl}d} show results when the RL agent prioritizes minimizing the index size. The heatmaps indicate that the trends for both workloads are similar, as the stored data index is identical. Larger block sizes consistently reduce the index size by decreasing the number of blocks and associated metadata. Among the index models, PLA with a maximum error of $256$ achieves the highest reward by reducing the number of splines, thus minimizing the index overhead. In both workloads, the chosen parameters achieve an estimated index size of less than $500MB$, which is half of the index size when the optimization is on the throughput. This is due to increasing the block size, which gives RocksDB a better compression ratio for the stored blocks, and also decrease the size of the LI model for each SST.



\begin{figure}[t]
  \centering
  \makebox[\columnwidth][c]{\includegraphics[width=0.9\columnwidth]{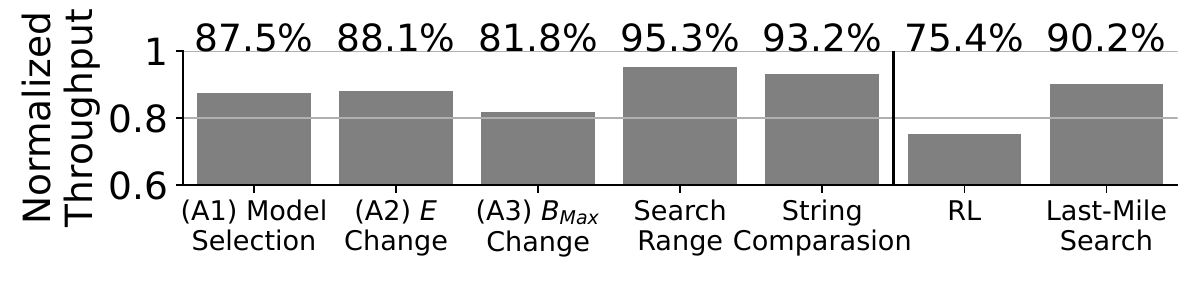}}
  \caption{\small{Ablation study. Bar plots show the normalized throughput to \texttt{DobLIX} full optimization.}}
  \label{fig:eval-ablation}
\end{figure}

\subsection{Ablation Study}
\label{sec:eval:ablation}
We perform an ablation study to evaluate the impact of optimizations in \texttt{DobLIX} under an RH workload (Fig.~\ref{fig:eval-ablation}). First, we tune the RL using a modified ZippyDB dataset with $16\text{KB}$ average values ($\text{SD}=5.2$) and find the optimal state (PRA model, $8\text{KB}$ $b_{max}$, and error range $E=8$). Then, we test these settings on the actual ZippyDB dataset while systematically disabling each optimization: (1) replacing RL-based model selection (PLA/PRA) with static PRA, (2) fixing $E=8$ instead of RL-tuning, (3) enforcing a static $16\text{KB}$ block size, (4) disabling last-mile search range optimization, and (5) removing string comparison optimization. The fully optimized \texttt{DobLIX} serves as the baseline. Results show that disabling RL-based optimizations ($A_1$--$A_3$) reduces throughput by $12.5\%$, $11.9\%$, and $11.2\%$, respectively. Removing search range optimization causes a $4.7\%$ drop due to inefficient searches, while disabling string comparison optimization lowers throughput by $6.8\%$, as keys often share prefixes that the optimization skips. We also disable whole RL optimization (1-3), and removing both last-mile optimizations (4, 5), which decrease the performance by $14.6\%$ and $9.8\%$, respectively.

\section{RELATED WORK}
\label{sec:related-work}

LSM read performance suffers from its layering; optimization focuses on Bloom filters, cache, and index. \textit{(1) Bloom Filter Optimization.}  Several works utilize filters to skip unnecessary reading since the filters do not return a false negative~\cite{li2019elasticbf,wang2024grf, sarkar2022dissecting}. However, the filters suffer from high false positive rates and excessive memory usage. \textit{(2) Cache Optimization.} LSbM-tree~\cite{teng2017lsbm} adds a buffer to minimize cache invalidations due to compactions. AC-Key~\cite{wu2020ac} introduces an adaptive caching algorithm to adjust the cache size based on the workload. Although caching accelerates the lookup of recently accessed data, it consumes significant storage as the cache size increases. Endure~\cite{Endure} optimizes the cache memory allocation adaptively based on the incoming workload. 
\textit{(3) Index Optimization.} SLM-DB~\cite{kaiyrakhmet2019slm} implemented a persistent global B+tree index on NVM, and Kvell~\cite{lepers2019kvell} adopts various memory index structures. DumpKV~\cite{DumpKV} trains a model on the KV lifetime to reduce the write amplification. Several studies~\cite{Bourbon2020,leaderkv2024,TridentKV2022} also built an LI using static data stored in SSTs to speed up read queries on SSTs. These studies improve indexing but overlook adjusting data access for persistent memory retrieval.

\section{DISCUSSION}
\label{sec:discussion}
\textbf{\texttt{DobLIX} Portability to various LSMs.}
All LSM variants (e.g., Cassandra, LevelDB) use lower-granularity units called blocks in their SST files to enable efficient data access.
Porting \texttt{DobLIX} to other LSM stores is simple: use the API for block size, then save/load the model after SST writes/reads. We show LI's portability by adapting Bourbon's LevelDB~\cite{leveldb} version to RocksDB, as discussed in \S\ref{sec:eval:baselines}, requiring minimal changes.

\noindent
\textbf{Mitigating RL Cold Start Problem.}
As shown in \S\ref{fig:eval-rl}, the DobLIX RL agent finds near-optimal parameters after 50 episodes (1,000 SST instances), even with just 32 states and pruned actions. This underlines RL's cold start problem~\cite{rl-coldstart}, but pre-trained models from similar workloads greatly cut training cost. For instance, transferring parameters from LOGN to AMZN (same RH point queries, both $\nu=0$ and $\nu=1$) requires only 5 fine-tuning episodes. Thus, pre-training transfers well across datasets for matching workloads, but switching workloads (e.g., point to range queries) on the same data still needs full retraining (30+ episodes).

\begin{figure}[t]
  \centering
  \makebox[\columnwidth][c]{\includegraphics[width=0.8\columnwidth]{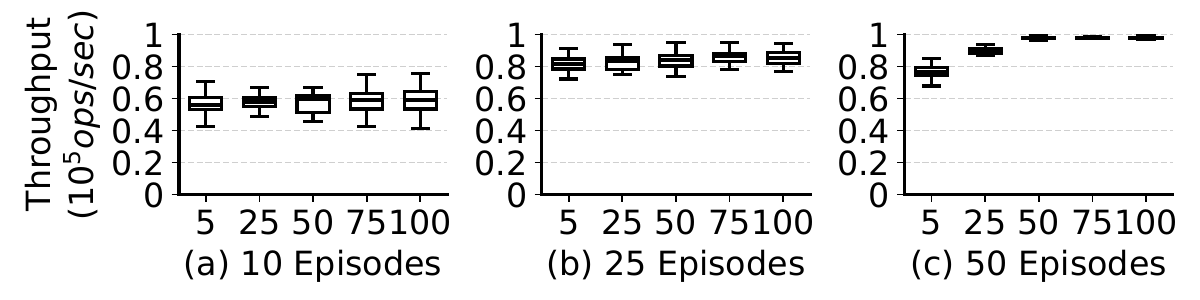}}
  \caption{\small{Final throughput robustness of the RL agent based on the percentage of initial samples to total data (x-axis), evaluated across three different total learning episodes (a, b, c). Each boxplot represents throughput distribution over 30 training runs, with the median (line), 1st quartile, and 3rd quartile (box) displayed.}}
  \label{fig:eval-robust}
\end{figure}

\noindent
\textbf{RL Agent Robustness.}
We test our RL agent’s robustness by varying initial training data ($5\%–100\%$) and training episodes ($10$, $25$, $50$). Results (Fig.~\ref{fig:eval-robust}) show: (1) Performance variance mainly depends on episode count, and (2) the agent is robust to small initial datasets. Notably, with only $25\%$ initial data and enough episodes, the agent reliably finds optimal solutions, indicating that our method works well even with limited data if training is sufficient.

\section{CONCLUSION}
\label{sec:conclusion}
We introduce \texttt{DobLIX}, an efficient LSM index that uses access-aware indexing on sorted SST keys during creation to enhance retrieval.
We also introduce an RL tuning agent in our method to optimize the LSM I/O block size and LI error range parameters in order to improve the trained LI performance.  
We show that \texttt{DobLIX} significantly improves read performance (up to $2.21\times$) in RocksDB, without compromising its write intensity. \texttt{DobLIX}'s adaptability to various workloads and data distributions allows it to be applied in a wider range of applications and scenarios. 

\noindent
\textbf{Future Work.}
We aim to extend our approach with multi-objective optimization for LSM parameter tuning (e.g., write amplification) during LI training. We also improve our RL agent by adding parameters, such as adjusting the Bloom filter’s false positive rate. Additionally, we investigate hybrid indexes that adaptively switch between learned and traditional types based on workload patterns for stable performance.

\clearpage
\nocite{*}
\renewcommand{\refname}{REFERENCES}
\renewcommand{\bibname}{REFERENCES}  
\bibliographystyle{ACM-Reference-Format}
\bibliography{sample}

\end{document}